\documentclass[a4paper,10pt]{article}
\usepackage{authblk}
\usepackage{geometry}
\usepackage{latexsym}
\usepackage{amsmath}
\usepackage{amsfonts}   
\usepackage{amssymb}
\usepackage{graphicx}
\usepackage{color}
\usepackage{parskip}
\usepackage[LGR,T1]{fontenc}
\usepackage{hyperref}
\usepackage{mathtools}
\usepackage{natbib}
\usepackage{tabularx}
\usepackage{subfig}
\usepackage{float}
\floatstyle{ruled} \restylefloat{figure}
\restylefloat{table}
\usepackage[toc,page]{appendix} \definecolor{red}{rgb}{0.8,0.0,0.0}
\hypersetup{colorlinks,breaklinks,linkcolor=blue,urlcolor=red,anchorcolor=red,citecolor=blue}
\usepackage{setspace}
\newcommand{\floatintro}[1]{
  \vspace*{0.1in}

  {\footnotesize
    #1
  }
  \vspace*{0.1in} }
\usepackage{appendix}

\title{Can one hear the shape of a target zone?} 

\author{Jean-Louis Arcand$^{1,2}$  \and
  Shekhar Hari Kumar$^1$ \and Max-Olivier Hongler$^3$\and Daniele Rinaldo$^4$\thanks{Corresponding author. \texttt{d.rinaldo@exeter.ac.uk} \\
The authors thank Joshua Aizenman, Ugo Panizza, Didier Sornette and C\'edric Tille for their precious insights. }
} \date{%
  $^1$The Graduate Institute of International and Development Studies, Geneva, Switzerland \\ \, $^2$Fondation pour les \`Etudes et Recherches sur le D\`eveloppement International (FERDI), Clermont-Ferrand, France\\ \ $^3$\'Ecole Polytechnique F\'ed\'erale, Lausanne, Switzerland \\ \, $^4$University of Exeter Business School, Exeter, United Kingdom}

\begin{document}

\maketitle
\vspace{0.3in}
\begin{abstract}
\small
 We develop an exchange rate target zone model with finite exit time and non-Gaussian tails. We show how the tails are a consequence of time-varying investor risk aversion, which generates mean-preserving spreads in the fundamental distribution. We solve explicitly for
  stationary and non-stationary exchange rate paths, and show how both depend continuously on the distance to the exit time and the target zone bands. This enables us to show how central bank intervention is endogenous to both the distance of the fundamental to the band and the underlying risk. We discuss how the feasibility of the target zone is shaped by the set horizon and the degree of underlying risk, and we determine a minimum time at which the required parity can be reached. We prove that increases in risk after a certain threshold can yield endogenous regime shifts where the ``honeymoon effects'' vanish and the target zone cannot be feasibly maintained. None of these results can be obtained by means of the standard Gaussian or affine models. Numerical simulations allow us to recover all the exchange rate densities established in the target zone literature. The generality of our framework has important policy implications for modern target zone arrangements. 
\end{abstract}

\begin{small}
\end{small}


\large

 \section{Introduction}\label{s:intro}

The exchange rate target zone literature pioneered by
\citet{krugman1991target} is based on a stochastic flexible price
monetary model in continuous time. This literature highlights the role
of market expectations concerning fundamentals in shaping exchange rate
movements. Given its assumptions of perfect credibility, it implies
that central bankers need only intervene marginally at the bounds of
the target zone or allow honeymoon effects to
automatically stabilize the exchange rate. The European Monetary
System (EMS) and the Exchange Rate Mechanism (ERM), which existed
from 1979 to 1999 (until participating countries adopted the Euro), provided a
natural test bed for this theory. The target zone model is both well accepted theoretically and has
provided the intellectual justification for a nominal anchor for
monetary policy. However, there is scant empirical support for the
validity of the framework. The U-shaped distribution within the target
band and the negative correlation between he exchange rate and the
interest rate differential implied by the canonical model have found
little counterpart in the data. In spite of this, the practice of
using target zones continues to this day, with the ERM-II target zone mechanism being an intermediate step to Euro adoption for new member states.\footnote{As of writing this paper, Croatia, Bulgaria and Denmark are in the ERM-II target-zone. Croatia and Bulgaria intend to adopt the Euro whereas Denmark has a special opt-out clause from Euro adoption.} It is likely that future new
member states will also participate in the ERM process, making target zone
modeling of current relevance. 

The seminal paper by \citet{krugman1991target} hinges on the
assumption of perfect credibility of the target zone, which gives rise
to a U-shaped distribution of the exchange rate. This implies that the
exchange rate spends most of its time near the bands of the zone, as
well as a negative relationship between the interest rate differential
and exchange rate volatility. Given this ``honeymoon effect'', the
central bank only has to intervene marginally at the bands. The only
source of risk in this model is the volatility of the Gaussian
distribution. The theoretical predictions of the model have been shown
not to hold empirically by \cite{mathieson1991}, \cite{meese1991} and
\cite{svensson1991}. This led to the development of so-called
second-generation models,
starting with \cite{bertola1992target} and \cite{bertola1993}.
Furthermore, \cite{tristani1994variable} and \cite{werner1995exchange} study
endogenous realignment risk, and include mean-reverting
fundamental dynamics.
\cite{dumas1992target}
and \cite{bessec2003mean}, using controlled diffusion processes, show
that the honeymoon effects are considerably weakened, putting into
question the necessity of a target zone when central banks intervene
intramarginally. 
\cite{serrat2000} generalizes the target zone framework to a multilateral setting, and shows how spillovers from third-country interventions can increase conditional volatilities compared to free-float regimes. 
\cite{bekaert1998target} and \cite{lundbergh2006time}
test the implications of the second-generation models, and find mixed
evidence with a slight tendency towards the intramarginal
interventions hypothesis. \cite{lin2008} proposes a framework with an interesting analogy to our model, where the spot rate can be stabilized by imposing a target zone on the forward rate. 
\citet{ajevskis2011target} extended the
basic target zone model to a finite termination time setting while
maintaining the assumptions of the original model: it is the closest
to our approach. \\


  Recently, \cite{studer2017swiss} and
\citeauthor{lera2016quantitative} (\citeyear{lera2016quantitative},
\citeyear{lera2018explicit} and \citeyear{lera2019currency}) find
empirical evidence for the target zone model for the EUR/CHF floor
target zone set by the Swiss National Bank between 2011 and 2015, the
latter mapping the Krugman model to the option chain. In particular, \cite{lera2015currency} show how the standard model can hold in specific cases, such as the EUR/CHF target zone,
because of a sustained pressure that continuously pushes the exchange
rate closer to the bounds of the target zone, which the central bank
tries to counteract. In this particular case, the sustained pressure
stemmed from the Swiss Franc being used as a safe asset in the middle
of the European crisis. 
\cite{rey2015dilemma} famously argued that the global financial cycles stemming from the United States generate additional risks for central banks targeting a nominal anchor. 
Additionally, \cite{gopinath2019} and \cite{kalemli2019} show how US monetary policy shocks can affect the exchange rate of a country with minimal USD exposure because of the dominant nature of the USD as a trade currency. 
This implies that the alignment process generated by a target zone naturally generates additional risk, which is radically different from the diffusive nature of idiosyncratic Gaussian noise. 
This risk destabilises the exchange rate fundamentals and creates an extra tendency to escape from its mean and move towards the boundary. 
All pre-existing attempts at modeling fundamental risk involve either the variance of Gaussian noise or the addition of ad-hoc jumps, or by assuming deviations from rational expectations. 

In this paper, we describe the dynamics of an exchange rate target zone with finite-time exit to a target currency, while accounting for the additional risk created by the convergence process for national central banks. We do so by assuming risk aversion for foreign investors to be subject to risk-on and risk-off shocks, which generate a country-risk premium in the uncovered interest rate parity condition. We show how this mechanisms introduces the dynamic equivalent of mean-preserving spreads in the fundamental process, which generate non-Gaussian tails in the exchange rate distribution.
These dynamics alter the country-risk premium of a small open economy, in which risk destabilises the fundamental process via a sudden bonanza or sudden stop of capital flows during the target zone process.
The nature of target zones with exit we describe require by construction a set final time at which both home exchange rate and fundamentals must match the ones of the anchor currency for successful adoption. One must therefore study both stationary (i.e. homogeneous) and non-stationary (time-dependent) exchange rate paths in order to analyze the dynamics of the target zone. 
In our framework we are able to solve analytically for both, allowing us to make four main contributions to the literature.

First, we show that due to exchange rate expectations in the fundamental exchange rate equation we are in the presence of "soft" boundaries. This implies central bank interventions are determined by the distance of the exchange rate to the bands as well as the external risk, i.e the tendency of the exchange rate to hit the bands. It turns out that the underlying dynamics are similar to the
phenomena famously described by \cite{kac1966}, where he asked whether one could \emph{``hear the shape of a drum.''}. In our framework the shape of a target zone can indeed be ``heard'', when the exchange rate is pushed to the sides of the target band by an additional external force: intuitively, this corresponds to the acoustic difference between striking a tense membrane (large shifts in risk aversion) versus a loose one. This is what we effectively describe in our paper: external risk is not simply subsumed in the variance of idiosyncratic fluctuations in the fundamental, but emerges as a destabilizing force that pushes probability from the center to the tails of the exchange rate distribution. This allows us to effectively endogenize the bands providing an intuitive explanation for the existence of smaller \textit{de facto} bands within larger \textit{de jure} target bands as well as the possibility of marginal and intra-marginal interventions in a target zone without assuming a specific foreign exchange intervention strategy. This mechanism has been shown empirically to exist and described heuristically \citep{lundbergh2006time,bessec2003mean}, but not yet formalized precisely.

Our second main contribution lies in the fact that 
we are able to characterize the minimum necessary time for which a target zone needs to be maintained for the home currency to successfully exit to the target currency. 
This is minimum time required for agents to "feel" the first effects of the home central bank’s actions aimed at reducing fluctuations of the exchange rate, compared to a free-float.
 It is also the minimum time necessary for agents to update their priors accurately, generating self-fulfilling expectations that create the honeymoon effect for future central bank actions. This result allows us to define precisely the feasibility of a target zone, something which has not been yet discussed in the literature. Feasibility corresponds to the central bank being able to reach the set central parity with the agreed bands at the chosen time horizon. Our model shows that considering non-stationary dynamics is paramount in determining whether the chosen horizon is feasible, and we characterize analytically the minimum required time necessary for the parity to be reached. Any smaller time horizon chosen by the central bank would not be feasible. In contrast, existing models assume away the problem by positing perfect feasibility and stationary dynamics.

Third, we show how large shocks to the investor risk aversion, leading to proportional increases in risk in the fundamental distribution, can potentially yield a regime shift once a certain risk threshold is crossed. This shift does not allow for honeymoon effects to happen anymore around the target zone bands, since the increase in risk destabilizes the exchange rate dynamics to the point that 
smooth-fitting procedures around the band cannot be applied by central bank interventions and the target zone becomes untenable. Beyond a critical threshold of risk, therefore, the target zone effectively cannot exist, and we show how this threshold is bounded below by the reciprocal of the target zone bandwidth. The mechanics behind this result show that the external risk destabilizes the fundamental and dominates its diffusive part. This implies that the central bank has to widen the target zone bands in order to maintain control of the exchange rate.  This result cannot be explained by standard models unless by explicitly incorporating \emph{ad hoc} endogenous devaluation risk. Our model, on the other hand, does not rely on the distribution of intervention or the level of reserves to generate this scenario, being instead a direct consequence of risk. 


Fourth, the standard case of exchange rate dynamics in a finite target zone with Gaussian-driven fundamentals is a simplified, limiting case of our model for which risk and variance are the same, which fails to provide a palatable explanation for well-known exits such as ERM-I. Correctly specifying risk aversion shocks leads to dynamics in which the exchange rate fundamental has a tendency to systematically escape its purely diffusive nature and move away from its expectation. 
As such, risk in our model emerges as a destabilizing force which runs counter to the best efforts of a central bank trying to maintain a target zone.
This causes persistent and potentially one-sided deviations from central parity. Moreover, we show that the effect of risk is both nonlinear and discontinuous. For low risk, our dynamics are similar to the standard model. As risk increases, the exchange rate process is increasingly destabilized and requires a monotonically increasing minimum time for the target zone exit to be reached successfully. 
We show how the model can fit a wide range of scenarios regarding feasibility and control, and we use Monte Carlo simulations to recover the different exchange rate densities presented by the established target zone literature. \\
Our paper has important policy implications. First, it establishes formally what is a feasible size for a target zone band, dependent on the level of underlying fundamental risk and the variance of the target currency. Second, it establishes the minimum time that a target zone must be maintained before any successful convergence can be achieved. Finally, it sheds light on the nature of external risk in target zone arrangements, and can inform central banks on mitigation strategies to limit exchange rate destabilization.\\

The paper is organised as follows. 
Section \ref{s:model} presents the model and its solution. Section \ref{s:gap}
discusses the connection between risk, target zone width and
feasibility, presents the emergence of regime
shifts once a critical threshold of risk is reached and shows the generality of our framework via numerical simulations. Section \ref{s:policy} discusses the policy implications of our model, while \ref{s:conclude} concludes and presents
an agenda for future research.

 \section{The model. Risk aversion shocks, mean-preserving spreads and exchange target zones with finite exit time}
\label{s:model}

In this paper we want to characterize a modern target zone mechanism in which the fundamental process can be destabilized by external risk factors, generating thick non-Gaussian tails in its distribution. Inclusion of these characteristics in the analysis is made necessary by the presence of risk-averse investors who have time varying risk-aversion modulated by the global financial cycle. Entering a target zone increases the capital market integration of the country in question which exposes countries' fundamentals to an increased share of global and regional risk factors.\footnote{\citet{fornaro2020finint} finds that entering a currency union increases financial integration between member states. This is due to reduction of currency risk and the associated easing of external borrowing constraints, driven in part by loss of national monetary and fiscal autonomy. A target zone setting is a \emph{quasi-currency union} with the chosen target zone band representing the range of expected fluctuations. Evidence from New Member States suggests that the magnitude of capital flows received may be very high even if the member state does not enter the target zone process for adopting the the Euro \citep{mitra2011_cflows}. This may be considered analogous to the index effects documented by \citet{hau2010index} for emerging market currencies.} In short, this framework allows us to consider additional fundamental risk arising from time-varying risk aversion generated by the global financial cycle when a currency enters a target zone.






 
The target zone framework depends critically on the uncovered interest rate parity (UIP) condition, with the currency in the target zone converging to the target nominal interest rate at time of exit to the currency union. The UIP condition requires risk-neutral preferences to hold: this is usually not the case when we are considering real-world situations, as investors are generally risk-averse. 
Risk aversion, however, is likely to change in time due to risk-on and risk-off shocks arising from global financial conditions. 
  Let us consider that investors face a standard problem of bond consumption with concave utility $U(c_t)$ discounted at $\gamma$. At any time $t$, if at the future time $t+1$ the agent's coefficient of relative risk aversion $-c U''/U' $ were to be incremented by an amount $ \lambda \in \mathbb{R}^+$ which can be either negative (risk-on) or positive (risk-off) with equal probability, yielding a new utility function $\bar{U}$.\footnote{This framework is equivalent to assuming heterogeneous investors, identical in everything except in risk aversion, where between $t$ and $t+1$ each changes her own risk aversion to a specific amount, and the resulting $\pm \lambda$ is the aggregate overall change in the representative utility function.} 
  This implies that the asset pricing kernel (the stochastic discount factor) will be given by 
 
 $$
\gamma \frac{\bar{U}'(c_{t+1})}{U'(c_t)} =  \gamma \frac{U'(c_{t+1})}{U'(c_t)} \Delta U'(c_{t+1}) = M_{t}\Delta U'(c_{t+1}) ,
 $$
 where $M_{t} $ is the pricing kernel without the change in risk aversion and $\Delta U'(c_{t+1})$ is the change in curvature of the utility function due to the change in risk aversion. Note that this last term is also a random variable. 
 More generally, if consumption of bonds is at two discrete time points but their evolution is continuous, this extra term is equivalent to the Radon-Nikod\'ym derivative for the change of measure between the densities generated by the differently curved utility functions. The investors' pricing kernel is therefore
 
 $$
\gamma \frac{\bar{U}'(c_{t+1})}{U'(c_t)} = \frac{d\mathbb{Q}}{d\mathbb{P}} \frac{d\mathbb{\tilde{Q}}}{d\mathbb{Q}}
 $$
 where $\mathbb{Q}$ is the foreign martingale measure of the home bond under the original measure $\mathbb{P}$, and  $\mathbb{\tilde{Q}}$ is the foreign martingale measure under the new utility function. The UIP condition is then given by 
 
 \begin{equation}
 \mathbb{E} \{ dX_t \}  \frac{(1+i^*_t)}{(1+i_t)} = \frac{d\mathbb{Q}}{d\mathbb{\tilde{Q}}},
\label{uipnew}
 \end{equation}
where $\mathbb{E} \{ d X_t \}$ is the expectation of the log exchange rate conditional on information available up to $t$ and $i^*$, $i$ are respectively the foreign and domestic interest rates. In \eqref{uipnew} the excess returns required to complete the no-arbitrage condition decrease with the investors' risk aversion, since $\frac{d\mathbb{Q}}{d\mathbb{\tilde{Q}}}$ increases with a realization of $+\lambda$ (decreased risk aversion) and vice versa. Equation \eqref{uipnew} is a modified UIP condition 
 where the time-varying risk premium is dependent on the change in investor risk aversion. If we assume again log-normality of the foreign bond, since the change in risk aversion is equally likely on each side (each $\pm \lambda$ is realized with probability 0.5), it's easily shown that that the new measure after the change in risk aversion is given by a Gaussian density identical to the pricing kernel without the curvature change, and an oscillating term that takes values $\pm \lambda$ with equal probability, represented by a Bernoulli variable.
 We note that the overall new measure $d \mathbb{\tilde{Q}}/d\mathbb{P}$ is still a martingale but is not Gaussian, even assuming an underlying (log) Normal distribution: the oscillation of the change in curvature of the utility function generates an extra term
 
 \begin{equation}
     \frac{d\mathbb{\tilde{Q}}}{d\mathbb{Q}} = \frac{1}{2}  \left (e^{-(x+\lambda)^2/2}+  e^{-(x-\lambda)^2/2} \right ),
     \label{pert}
 \end{equation}
which is exactly the perturbation of a Gaussian process by means of a Bernoulli variable in the drift. We therefore have a risk premium that is dependent on the oscillation of investors' risk aversion, $\pm \lambda$ with equal probability. \\

Let us now include \eqref{uipnew} and \eqref{pert} in the exchange rate dynamics.
 %
 As standard procedure in the literature, the fundamental process for the exchange rate $f_t$ evolves according to $df_t = d v_t + d m_t$, where $v_t $ is a money demand shock (velocity) $m_t$ is money supply, usually assumed to be controlled by the central bank.
As shown in Appendix \ref{monetary}, the varying risk premia from the modified UIP condition \eqref{uipnew} can be included in the velocity in a monetary model of exchange rate determination, and
 we can write the fundamental process 
 as the stochastic differential equation



\begin{equation}
  df_t=  \lambda \mathcal{B} dt  + \sigma d W_t, \qquad f_{t=0} = f_0, \label{dmps}
\end{equation}

\noindent in the probability space $(\mathbb{R}, \mathcal{F}, P)$ where $dW_t$ is the standard Brownian motion and $\mathcal{B}$ is a Bernoulli random variable obtaining values $\{-1, 1 \} $ each with probability 0.5 and
$\lambda \in \mathbb{R}^+ $. Without any loss of generality, \eqref{dmps} can be rescaled as: 

\begin{equation}
  df_t= \beta \mathcal{B}dt  + d W_t,  \label{dmps2}
\end{equation} 
where $\beta = \lambda/ \sigma^2$ is the rescaled risk parameter. 
The stochastic process \eqref{dmps2} driving the fundamental is the dynamic equivalent of a mean-preserving spread, and has been studied by \cite{arcand2018increasing}. 
Risk aversion shocks in the velocity, therefore, cause an increase in risk in the fundamental that push probability away from the mean and generate non-Gaussian tails, whilst leaving the systematic average unchanged. 
As seen in Eq. (\ref{uipnew}) and \eqref{pert}, such shocks cannot be represented by Gaussian fluctuations, and we prove this formally in Appendix \ref{assumptions}.
This also allows us to precisely characterize the interplay of diffusive fluctuations (variance) and destabilizing forces\footnote{The term $\beta \mathcal{B}$ is indeed a force, being the derivative of the probabilistic potential of the process $f_t$.} (risk, via changes in investor risk aversion):
the tendency of external risk to shift the exchange rate
away from the mean and towards the bounds of the zone is counteracted by the central bank's efforts to maintain the fundamental fluctuating around its mean. This is precisely what is observed by \cite{lera2015currency}. \\

By definition of a target zone, the central bank requires the fundamental process to remain bounded within a set interval $[\underline{f}, \overline{f} ] \in \mathbb{R}$. By adjusting money supply, the central bank regulates the fundamental process within this interval, also known as the target band. As shown in Appendix \ref{monetary}, the equation for the log-exchange rate $X_t$ under the modified UIP condition \eqref{uipnew} in the interval $[\underline{f}, \overline{f}] $ can therefore be written as the regulated stochastic differential equation %
\begin{equation}
  \label{EXCHA}
  X_t =  f_t + \frac{1}{\alpha} \mathbb{E}\left\{ d X_t \right\}  \qquad f_t  \in [\underline{f}, \overline{f}] \ \ \forall t\in [0,T].
\end{equation}
%
where $f_t$ evolves according to \eqref{dmps2}, equipped with the reflecting (``smooth pasting'') boundary conditions

\begin{equation}
\partial_f X_t \mid_{f = \underline{f}} \,\, =\,\;  \partial_f X_t \mid_{f= \overline{f} }\,\,   =\,\; 0. \label{refl}
\end{equation}
At a fixed time $T$ the spot exchange rate is set to exit the target zone and match the target fundamentals. We are interested in the exchange rate dynamics generated by \eqref{EXCHA} throughout the time interval $[0,T]$, and therefore explicitly allow time-dependent dynamics $X_t = X(t,f_t)$ and study non- stationary behavior. The term $1/\alpha$, with $0 < \alpha < 1$ is the absolute value of the semi-elasticity of money with the nominal interest rate. As this quantity is always greater than unity, we interpret as a frequency (i.e.
$1/[{\rm time\,\, unit}]$) which modulates the size of the forward-looking time window. For simplicity of exposition, we focus our  attention to target zones symmetric around 0 of the form $[-\overline{f}, \overline{f} ] $, although all our results hold for general bounds. 
We can now state the main result of the paper.\\
\noindent \textbf{Proposition 1:} \emph{The solution of the stochastic differential equation} \eqref{EXCHA} \emph{reflected via} \eqref{refl} \emph{in the bounded domain $[0,T] \times [\underline{f}, \overline{f} ] \to \mathbb{R}$ is given by the process}

\begin{equation}
    X(t, f) = X^{*}(t ,f) + X_S(f),
\end{equation}
\emph{which is the sum of stationary and non-stationary solutions. The stationary solution is given by }
\begin{equation}
  \label{XPH}
  X_S(f) = \cosh(\beta f)^{-1} \left[  A {\cal Y}_1 (f)+ B  {\cal Y}_2 (f)+{ \cal Y}_{P}(f)\right ],
\end{equation}
\emph{where we have:}
\begin{eqnarray*}
      {\cal Y}_1(f) &=& \exp\left ( +\sqrt{ \left[ \beta^{2}   +  {2\alpha  }\right]} f \right) , \\ 
      {\cal Y}_2 (f) &=& \exp \left ( -\sqrt{ \left[ \beta^{2} +  {2\alpha   }\right]}f \right )  , \\ 
      {\cal Y}_P (f)  &=& \frac{\left [  \alpha f  \cosh (\beta f) +
    \beta  \sinh (\beta f )\right] }{ \alpha}
\end{eqnarray*}
\emph{and the constants $A,B$ are obtained by smooth-fitting at the boundaries}

\begin{equation}
  \label{SM}
  \partial_f X_S(f) \mid_{f = \underline{f}} \,\, =\,\;  \partial_f X_S(f) \mid_{f= \overline{f} }\,\,   =\,\; 0.
\end{equation}
\emph{The non-stationary solution is obtained by means of an eigenfunction expansion and is given by}

\begin{equation}
  X^{*}(t,f)  =  \cosh(\beta f)^{-1} \sum_{k=1}^{\infty} c_k \exp {\left[-(  \Omega^{2}_k  + \rho)   (T-t)  \right]} \sin\left ( \sqrt{2} \Omega_k f\right ) 
            \label{EQTRANSIENT}
\end{equation}
\emph{with} $\rho = {\beta^{2} \over 2} + \alpha $ \emph{and Fourier coefficients given by } 
$$
  c_k =  -{1 \over \overline{f}} \displaystyle\int_{- \overline{f}} ^{+\overline{f}}X_S(f) \sin\left ( \sqrt{2} \Omega_k f \right ) df.
$$
\emph{This solution is defined over a complete set of real eigenvalues $\{ \Omega_1, \dots, \Omega_k,\dots \}, k = \mathbb{N}^+$ that solve }
\begin{equation}
\sqrt{2} \Omega_k \cot \left (  \Omega_k \overline{f} \right )
- \beta  \tanh( \beta \overline{f} )  = 0, \qquad \forall k \in \mathbb{N}^+ \label{eigenv}
\end{equation}
\emph{and span the discrete spectrum }$  \Omega := \left\{ \Omega_k (\beta , \overline{f})\right\}.$ \\
\\
\textbf{Proof:} See Appendix \ref{derivations}. \\

\begin{figure}[htbp]
  \caption{Effect of varying $\beta$ on stationary exchange rate
    dynamics}\label{fig:betas}
  \begin{center}
    \includegraphics[scale=0.7]{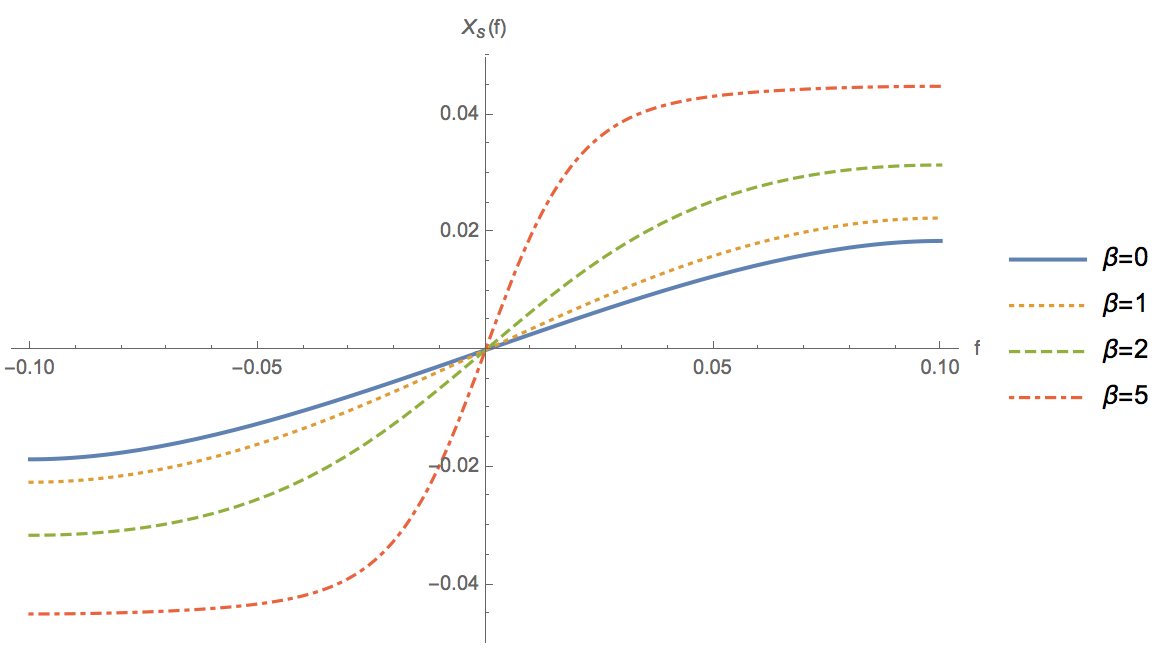}
  \end{center}
  \label{stat}
\end{figure}

 To see how the results are unaffected by both the band choice and its symmetry around 0, notice that the bounds enter the particular solution only via the scalar quantities $A$ and $B$, and in the general solution via the integration limits of the Fourier coefficients. An illustration of the stationary solution \eqref{XPH} is
presented in Figure \ref{stat}, which also shows how an increase in
the riskiness $\beta$ of the fundamental prompts the (stationary)
exchange rate to behave more independently of the dynamics of the
fundamental. At high levels of $\beta$, the exchange rate dynamics are
driven mostly by the risk and depend less on fundamentals, especially
around the bounds, as represented by the steepening of the central
slope. In this figure, $\overline{f}=10\%$ and we assume a quasi-daily
time step for the expectation $\alpha =0.8$. Our parametrization of $\alpha = 0.8$ corresponds to a case of fast agent updating, which is similar to the case studied by
\citet{ferreira2019expectation} and
\citet{coibion2015information}. Changing the $\alpha$ to a lower fundamental updating frequency will reduce the sensitivity of the exchange rate to the fundamentals.

\begin{figure}[htbp]
   \caption{Target band and spectrum}\label{fig:betas_spectrum}
  \begin{tabular*}{\textwidth}{@{} c @{\extracolsep{\fill}} c @{}}
    \includegraphics[scale=0.5]{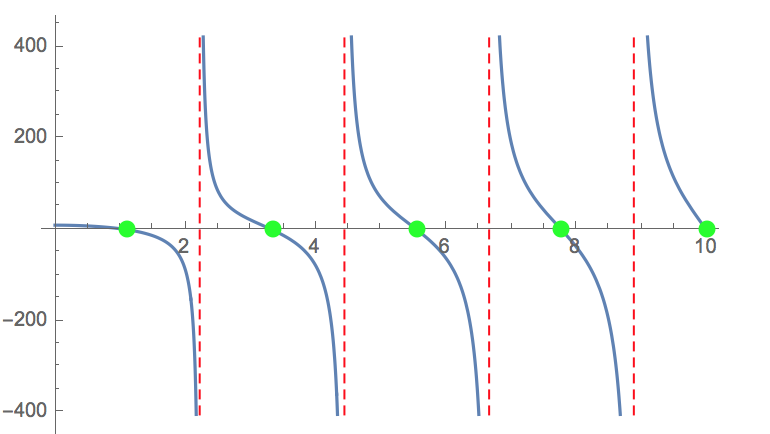} & 
    \includegraphics[scale=0.5]{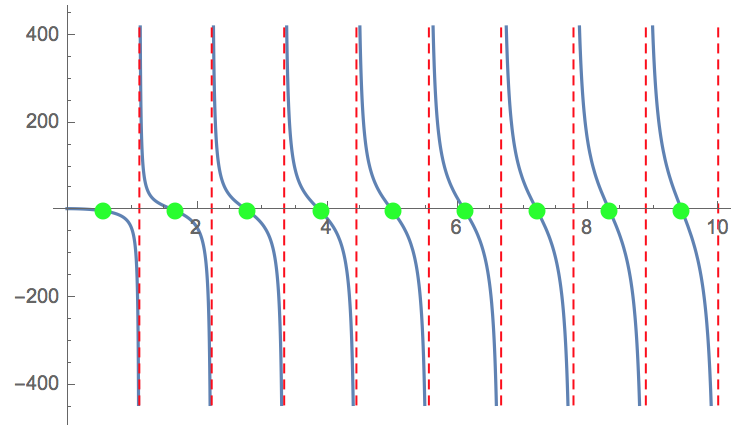} \\
    (a) $\bar{f}=10\%$ & (b) $\bar{f}=20\%$ \\
  \end{tabular*}
 \floatintro{Graphical illustration of the solution of
      equation \eqref{eigenv}, showing the effect of varying $\bar{f}$
      on the spectrum $\Omega_k$.}
  \label{spec}
\end{figure}
\begin{figure}[htbp]
  \caption{Non-stationary dynamics}\label{fig:betas_tau}
  \begin{tabular*}{\textwidth}{@{} c @{\extracolsep{\fill}}}
    \includegraphics[scale=0.55]{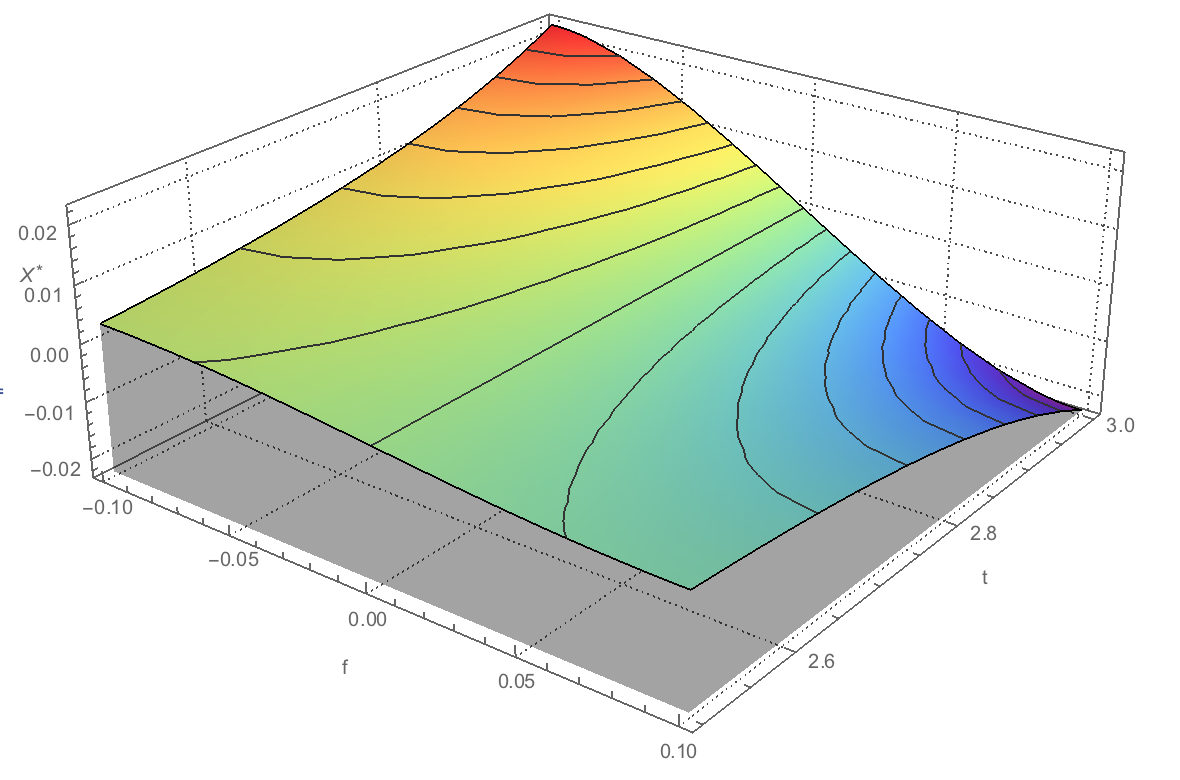} \\
    (A) \\
    \includegraphics[scale=0.55]{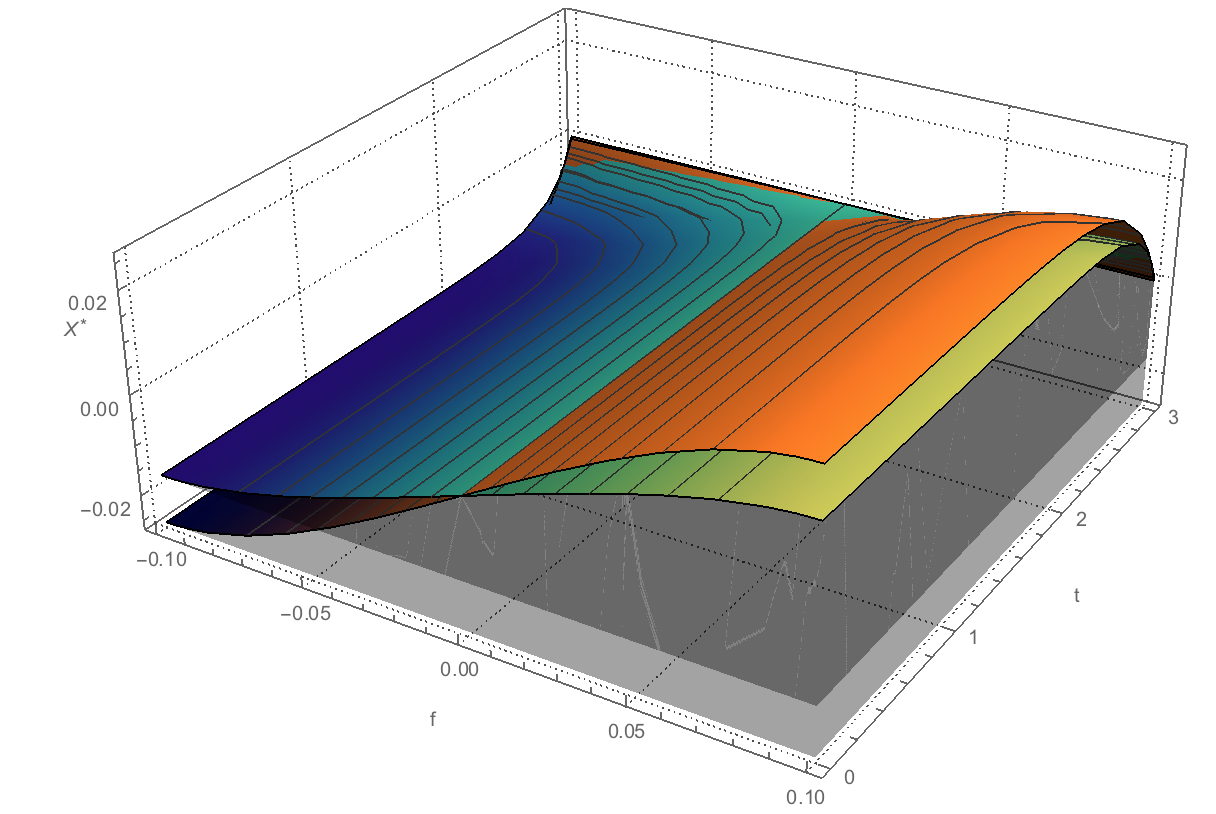} \\
    (B) \\
  \end{tabular*}
 \floatintro{This figure shows the
      evolution of $X(T-t,f)$ of the non-stationary dynamics in the target
      zone. Panel (A) shows the behavior of the time-dependent part: we assume a target zone which has been set to 
      $T=3$ years, with $\beta=1$ for a given set of fundamentals. For
      the sake of brevity we truncate the figure towards the end of
      the target zone to effectively illustrate the non-stationary
      dynamics. Panel (B) shows the full dynamics for an increase in risk.  Here we have assumed a
      target band symmetric around zero, i.e.
      $\bar{f} = 10\% = - \underline{f} $. We also assume $\alpha = 0.8$. We truncate the eigenfunction expansion
      at 50. The second panel illustrates the change in dynamics from
      $\beta=0$ (Gaussian) to $\beta=5$.}
  \label{transient_tab}
\end{figure}
%



\noindent For any $k \in \mathbb{N}^+$, the corresponding
$\Omega_k(\beta , \overline{f})$ solves \eqref{eigenv}, and has to be calculated numerically. For a general
$\beta > 0$, one observes that the successive eigenvalues are not
evenly spaced, and display a distance which decreases in $k$. The spectrum is
controlled by the width of the target zone $\bar{f}$: the wider the
band, the smaller the separation. The spectrum and its relationship
with the target band size are illustrated in Figure
\ref{spec}. Observe also that in the limit $\beta =0$, one
straightforwardly verifies that from Eq.(\ref{eigenv}) one obtains the
evenly spaced set
$\Omega_k(0, \overline{f}) = (2k+1){\pi \over 2 \overline{f}}$.




When $t=T$, from Eq.(\ref{EQTRANSIENT}), by
construction of the Fourier coefficients $c_k$, we have
$X^{*}(T,f)= - X_S(f)$ and so $X(T,f)= X^{*} ( T,f) + X_S(f) =0$ thus
reaching the required fixed parity. An illustration of the non-stationary
exchange rate dynamics, as well as the overall transition dynamics
throughout the time interval $[0,T]$, is presented in Figure
\ref{transient_tab}.  The solution allows one to express the movements of
the exchange rate via a weighted sum of its stationary behavior, its
distance to the exit time and the distance between its value at any
time $t$ and the target band. The eigenvalues modulate the frequency
of both fundamental and exchange rate movements within the band. The
Fourier coefficients $c_k$ represent the impact of the size of the
target band in the overall dynamics, via their weight on the infinite
series of frequency components (the ``harmonics'' of the exchange rate
path). Loosely speaking, this formulation of the solution allows one
to describe the sensitivity of the exchange rate to the distance to
the target band. Once the eigenvalues and the eigenfunctions are
known, as famously asked by \cite{kac1966}, \emph{``if one had perfect
  pitch''}, one would be able to ``hear'' the shape of the target
zone. Indeed, the time-independent part of the problem is a one-dimensional Neumann problem on the boundary
  $\partial D = [\underline{f}, \overline{f}]$
$$
\begin{cases}
  \Delta f + \Omega f = 0 \\
  \nabla f |_{\partial D} = 0,
\end{cases}
$$
which is exactly the problem of finding the overtones on a vibrating
surface. \\

This formulation of the solution allows us to uncover the unique
nature of the smooth-pasting conditions: the exchange rate process is
not reflected at the bounds in the probabilistic sense, since this
would have been modeled as a zero derivative condition on the
transition probability density function. The eigenfunction expansion shows that in a target zone there exist ``soft'' boundaries, where the central bank interventions are determined by the interplay of the distance of the exchange rate to the bounds as well as the tendency of the fundamental to hit them (the risk). 
This allows us to ``endogenize'' the bands: because of $\mathbb{E} \{ dX_t \}$ in the exchange rate equation \eqref{EXCHA}, we
have a second-order term which allows us to solve the equation in its
Sturm-Liouville form. The Fourier
coefficients in Proposition 2 modulate the sensitivity of the exchange rate to the
distance to the band, allowing for the central bank to intervene
whenever the fundamental is felt to be approaching the
bounds. This mechanism is a direct translation of how much
the fundamental tends to escape and how much the central bank needs to intervene marginally or intramarginally: it is a direct consequence of the presence of expectations in the exchange rate equation.
In other words, the higher the tendency to hit the bounds, the greater is the likelihood that the central bank will actually intervene
intramarginally, with increasingly less weight placed on the actual
position of the fundamental within the band. 
There exist therefore both \emph{de jure} and
\emph{de facto} bands, which is a feature of target zones observed empirically by \cite{lundbergh2006time}: if the
\emph{de jure} band is large, expectations over the magnitude of risk react to a narrower \emph{de facto} band. This is a phenomenon
commonly observed in most ERM countries but not yet formalized\footnote{See Figure 2 in \cite{crespo05}}. 



 \section{Spectral gap, target zone feasibility, regime shifts and numerical simulations}
  \label{s:gap}

In this section we discuss the key contributions of our framework: the role of the spectral gap in determining target zone feasibility, the characterization of the minimum feasible time to exit via the relaxation time of the exchange rate process, and how the fundamental risk can generate regime shifts. Furthermore, via numerical simulations we show how one can recreate via our model the various exchange rate densities described in the established literature. \\

Let us begin by studying the interplay between the risk parameter $\beta$, the
size of the target band $[-\overline{f}, +\overline{f}$] and the time horizon $T$ at which to reach the target
zone. We first note that at the initial time $t=0$, from
Eq.\eqref{EQTRANSIENT} we have $X^{*}(0,f)\approx 0$ and therefore
$X(0,f)= X^{*} (0,f) + X_S(f) \approx X_S(f)$. Since
$\Omega_1(\beta , \overline{f})< \Omega_2 (\beta , \overline{f})<
\cdots $, one can approximately write:

$$
\begin{array}{l} 
  X(T, f) \simeq X_S(f)  + {\cal O}\left( e^{-( \Omega^{2}_{1}  + \rho)  T} \right) .
  \qquad \qquad \qquad 
\end{array}
$$
While for the exact solution we should have $X(T, f) = X_S(f) $, one
sees immediately that $ X(T-t, f)= X_S(f) + X^{*}(T-t ,f) $ with
$X^{*}(T-t ,f) $ given by Eq.\eqref{EQTRANSIENT} nearly matches the
exact solution, provided we have an horizon interval
$T\gtrsim t_{\rm relax} $ where
$t _{\rm relax}:= \left( \Omega^{2}_{1} + \rho\right)^{-1}$ is the
characteristic relaxation time of the exchange rate process. This
provides a validity range for the non-stationary dynamics given
by the expansion Eq.(\ref{EQTRANSIENT}).

Hence, at time $t=0$, the required initial probability law $X_S(f) $ is reached only for a large enough time horizon
$T \gtrsim t_{\rm relax}$. This now enables us to link the non-stationary
dynamics of $X^*(t,f)$ to the \textbf{feasibility} of the target zone:
the relaxation time $\tau_{\rm relax}$ determines the \emph{minimum
  time interval} for which a feasible target zone may be
maintained. The larger $\beta$ (the risk of the fundamental, stemming from larger shifts in agents' risk aversion), the
greater is the tendency of the fundamental to escape from its mean;
the authorities need therefore to maintain the target zone for a
longer minimum duration. An increase in risk, for a given $\bar{f}$,
implies that the target zone would have to be set for a longer horizon
$T$ to be feasible. Alternatively, for a given risk $\beta$, an
increase of the target zone width $\overline{f}$, requires a longer
minimal $T$ implementation to ensure the overall feasibility of the
policy. In other words, the central bank has to impose that the time
horizon $T$ is at least as large as the relaxation time
$ t_{{\rm relax}}$.

An intuitive interpretation of the relaxation time in this framework
is to understand $t_{relax}$ as the characteristic elapsed time
required to ``feel'' the first effects of the home central bank's
actions aimed at reducing fluctuations of the exchange rate, compared
to a free float. The bank's actions may be then viewed as a \emph{de
  facto} reduction of the target zone band over time, whilst the
\emph{de jure} band remains unchanged. Furthermore, $t_{relax}$ can be interpreted as be the minimum time for agents to update their
priors accurately, generating self-fulfilling expectations that create the honeymoon effect.

\begin{figure}[htbp]
  \caption{Interaction between $\beta$ and
    $\overline{f}$}\label{fig:betas_widths}
  \begin{tabular*}{\textwidth}{@{} c @{\extracolsep{\fill}}}
    \includegraphics[scale=0.8]{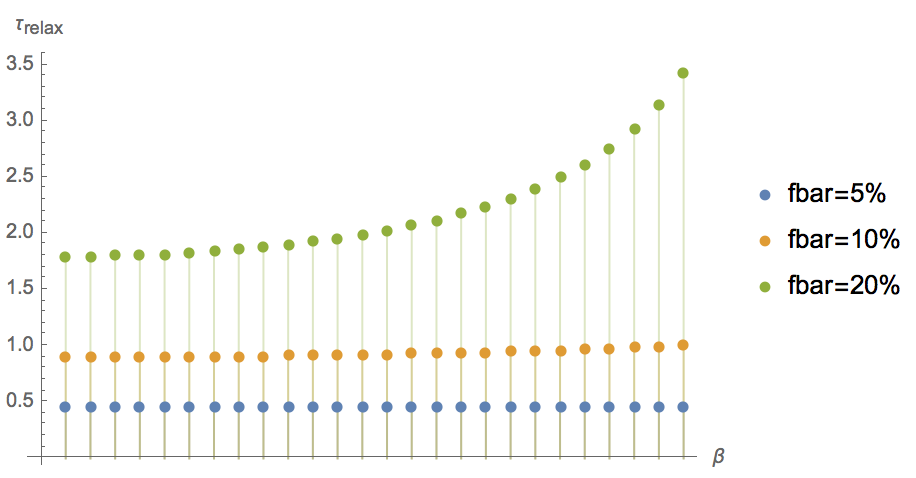}\\
  \end{tabular*}
  \floatintro{\textit{Note:} This figure shows the
      interaction of varying risk ($\beta$) and varying the band size
      ($\overline{f}$). An increase in risk, for a given
      $\overline{f}$, implies that the lowest eigenvalue $\Omega_1$
      falls (Panel (A)). The inverse of this value controls the
      $t_{{\rm relax}}$.}
  \label{trelax}
\end{figure}

The inverse of the relaxation time is determined by the
\textbf{spectral gap}, which is the distance between 0 and the
smallest eigenvalue. We therefore have the relationship
$(t_{{\rm relax}})^{-1} = (\Omega^{2}_1 + \rho)$. The spectral gap
controls the asymptotic time behaviour of the expansion given by
\eqref{EQTRANSIENT}, and it is continuously dependent on risk $\beta$
and band $\bar{f}$. This relationship is illustrated in Figure
\ref{trelax}.
%
%

Let us now study analytically the behaviour of the solution $\Omega_1$
of the transcendental Eq.\eqref{eigenv}. Writing
$z= \sqrt{2}\Omega_1 \overline{f}$, Eq.\eqref{eigenv} implies
%
%
%
%
%
that the product $\beta \overline{f}$ is the determinant of the
amplitude of $\Omega_1$. One can therefore immediately conclude that two limiting situations can be reached:

$$
\left\{
  \begin{array}{l}
    \beta \overline{f} <<1 \quad \Rightarrow \quad   \,z \lesssim {\pi \over 2}  \quad \Rightarrow   \quad \Omega_1 \lesssim  {\pi \over 2 \overline{f}}  \quad \,  \Rightarrow \quad   t_{{\rm relax}}^{-1} \lesssim \left[ {\pi \over 2 \overline{f}} \right]^{2}+ {\beta^{2} \over 2}  + \alpha , \\ \\
 
    \beta \overline{f}>>1 \quad \Rightarrow \quad   z \gtrsim{ \pi }  \,\quad \Rightarrow  \quad  \Omega_1 \gtrsim  {\pi \over  \overline{f}} \, \quad\,  \Rightarrow \quad t_{{\rm relax}}^{-1}  \,\gtrsim \,\left[{ \pi  \over   \overline{f} }\right]^{2}  +  {\beta^{2} \over 2}  +  \alpha.
  \end{array} \label{spectral}
\right.
$$

\noindent and therefore:

\begin{equation}
  \frac{1}{\left[{ \pi  \over   \overline{f} }\right]^{2}  +  {\beta^{2} \over 2}  +  \alpha } \leq t_{{\rm relax}} \leq \frac{1}{ \left[ {\pi \over 2 \overline{f}} \right]^{2}+ {\beta^{2} \over 2}  +  \alpha } .  \label{approx_bounds}
\end{equation}
Eq.(\ref{approx_bounds}),  together with Figure \ref{trelax} shows
how an increase in risk $\beta$ affects $t_{\rm relax}$ more strongly
when the exchange rate is allowed to float in a wider band width
$\bar{f}$.


\begin{figure}[htbp]
  \caption{Risk, target band and regime shifts}\label{regimeshift}
  \begin{tabular*}{\textwidth}{@{} c @{\extracolsep{\fill}} c @{}}
    \includegraphics[width=7cm]{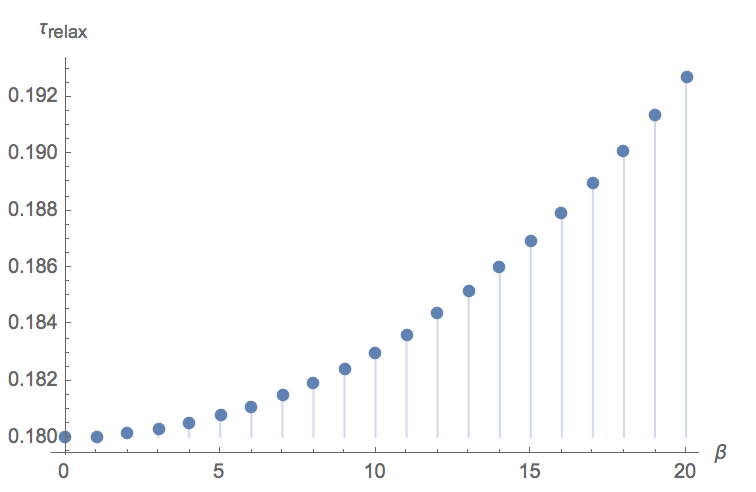} & \includegraphics[width=7cm]{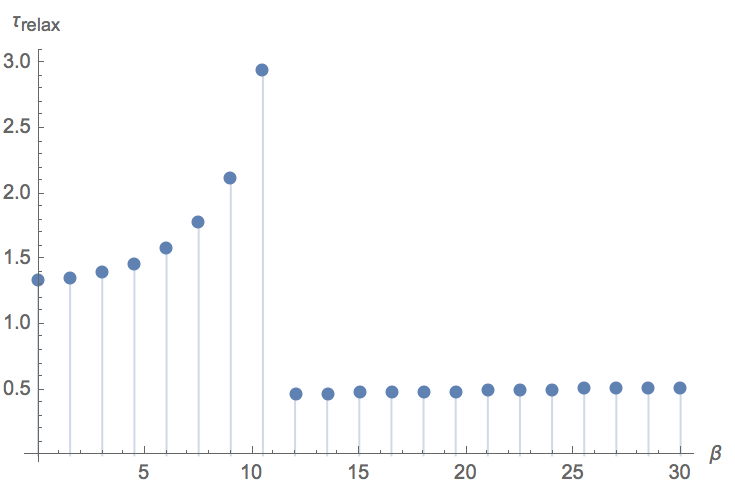} \\
    (a) $\bar{f}=2\%$ & (b) $\bar{f}=15\%$ \\
  \end{tabular*}
  \floatintro{Regime shift and eigenvalue jump as a
      function of risk, for different target bands}
  \label{spec}
\end{figure}

\noindent
An unique phenomenon that emerges in
our framework, is the emergence of a
\emph{regime shift}. Figure \ref{spec}(b) shows that for a
large enough target band, after a threshold level in $\beta$, the
relaxation time suddenly jumps to a much lower value and remains
almost constant (though very slowly increasing) for further increases
in risk. This effect happens because when the tendency $\beta$ of the
noise source driving the fundamental reaches and surpasses a certain
level, the destabilizing risk component in the noise source overcomes the
diffusion. The force $\beta \mathcal{B} $ in the mean-preserving
spread becomes the main driver of the stochastic process driving the
fundamental, and therefore $f_t$ becomes a process
with a tendency to escape from its mean that is stronger than the
tendency to diffuse around its central value. While this may look like
a sudden emergence of supercredibility, it is in fact the opposite:
the target zone cannot be feasibly held, as the fundamental process escapes its initial position with such force that it hits the band at every $dt$, and interventions need to be almost continuous. The central bank will have to either increase the size of the band or to allow the spot rate to float freely. This has a direct
implication for honeymoon effects, as shown in the following Proposition:\\

\textbf{Proposition 2. Risk, regime shifts and honeymoon effects.}
\emph{There exists a threshold level of risk $\beta^e \in \mathbb{R}^+$ which generates a regime shift. This is caused by a jump in the spectrum $\Omega$, as the elasticity with respect to the fundamental process of each eigenfunction $\psi_k$  associated to the eigenvalue $\Omega_k$ must always match the underlying mean-preserving probability spread:}

\begin{equation}
  \frac{\partial_f \psi_k (\Omega_k, f_t)}{\psi_k (\Omega_k, f_t)} =  \text{MPS}(\beta, f_t) \qquad \forall \ \Omega_k( \beta, \overline{f}) \in \Omega   \label{eigelast}.
\end{equation}  
\emph{For $\beta \geq  \beta^e$, the smooth-fitting procedure at the
boundaries cannot be applied and honeymoon effects when the fundamental approaches the band become
unobtainable.}\\

\textbf{Proof}: See Appendix \ref{honeymoon}. \\

\begin{figure}[htbp]
  \caption{Risk threshold, distance to the bands and honeymoon
    effects}
  \begin{tabular*}{\textwidth}{@{} c @{\extracolsep{\fill}} c @{}}
    \includegraphics[width=8cm]{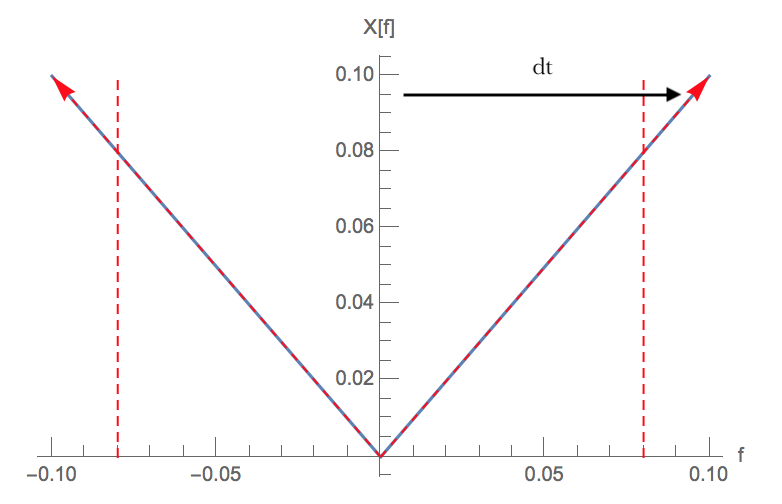} & \includegraphics[width=8cm]{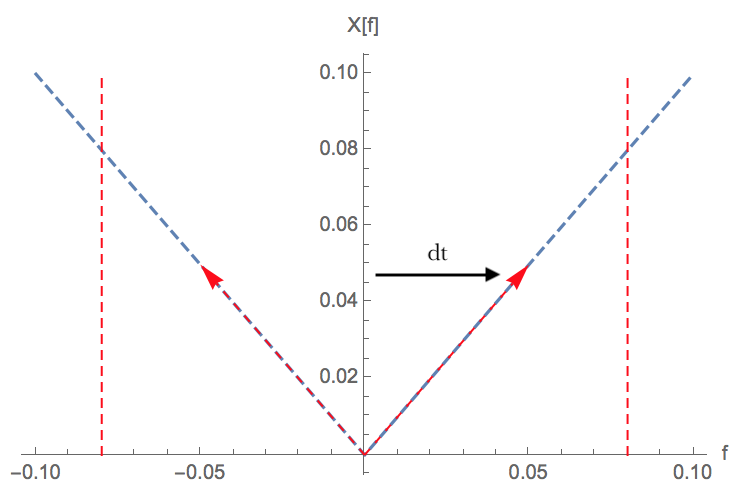} \\
    (a) $\beta > \beta^e$. No honeymoon effects & (b) $\beta < \beta^e$.\\
  \end{tabular*}
  \floatintro{Large risk shocks vs. diffusion-driven
      regimes}
  \label{dtbands}
\end{figure}

\begin{figure}[htbp]
  \caption{Risk and eigenvalue jump}\label{jump}
  \begin{center}
    \includegraphics[scale=0.7]{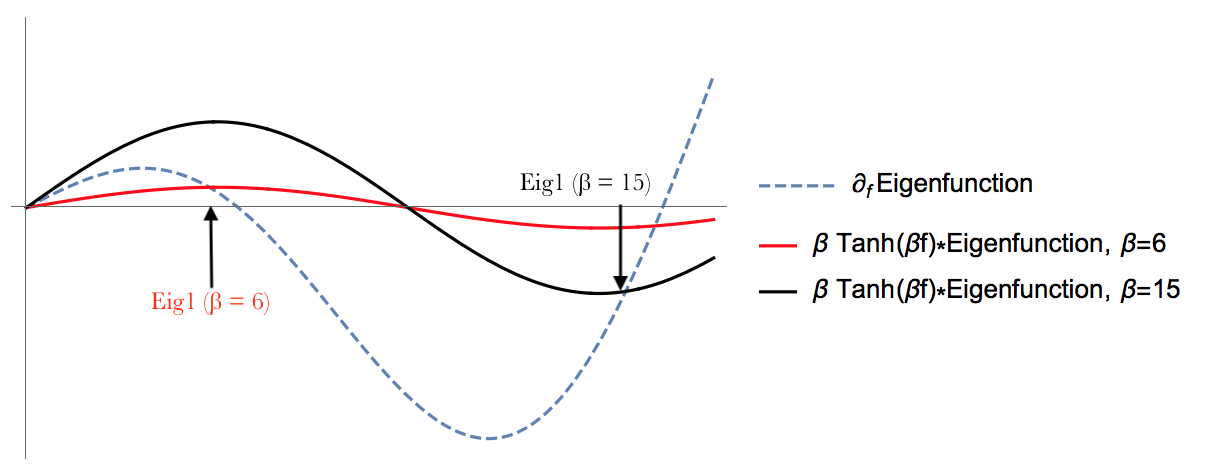}
  \end{center}
  \floatintro{\textit{Note:} Regime change For $\beta=15$.
      The force $\beta \tanh(\beta f)$ (black curve)
      overcomes the diffusion component and generates the first
      eigenvalue jump. For $\beta=6$ (red curve), the regime has not
      yet shifted. Here $\bar{f} = 0.1$, $\sigma = 1, \alpha=0.8$. }
\end{figure}

Proposition 2 implies that a high level of risk
denies a central bank monetary autonomy up until the moment of
entering the currency zone. This phenomenon is illustrated in Figure
\ref{dtbands}.
The first term in \eqref{eigelast} is a total sensitivity term, closely related to the elasticity of the eigenfunction with respect to the fundamental, and it represents the overall variation of the exchange rate with the fundamental. The second term represents the increase in risk, as well as the destabilizing component that represents the tendency of the fundamental to hit the target bands. The solution of this equation yields the spectrum $\{ \Omega_k \}$, for $k=\mathbb{N}^+$. The difference of the two terms represents the residual tendency of the home country fundamental to avoid converging to the target fundamental. The spectral gap, therefore, represents the intensity of the probability spread. The regime shift will happen at a threshold value $\beta^e$, only obtainable numerically, for which the spectral gap will suddenly jump upwards: the destabilizing force has dominated over the diffusive part and the first eigenvalue jumps higher. The oscillating part of the expansion increases in frequency, and the time-dependent exponential decay increases in speed. A graphical illustration is shown in Figure \ref{jump}: one can easily show that the lower bound for the threshold $\beta^e$ is given by $1/\bar{f}$. This allows one to uncover the close relationship between the regime shift and the size of the target band. This regime shift \textbf{cannot} occur with a Gaussian process or with mean-reverting dynamics. \\
%
%

In the diffusion-driven regime (characterised by a relatively low
$\beta < 1/ \overline{f}$), one observes that an increase of risk
implies a decrease in sensitivity, since $t_{relax}$ is
increasing. This may seem counterintuitive: but it must be remembered
that at time $t=0$, the initial condition is the stationary solution
of the central bank-controlled diffusion for the given
risk. Increasing $\beta$, therefore, is likely to load the stationary
probability mass accumulated in the vicinity of  the target zone
boundaries. Escape from this stationary state by bank action becomes
 more difficult, ultimately leading to an increase of
$t_{relax}$. Conversely, in high risk regimes where
$\beta > 1/ \overline{f}$ and where the destabilizing dynamics dominate, the
boundaries of the target zone are systematically hit by the
fundamental. In this situation, the central bank will intervene almost
entirely intramarginally regardless of whether the fundamental is
actually close to the bands, since honeymoon effects cannot exist
anymore. This allows, in Eq.(12), for a sudden
reduction of the probability mass located at the bounds, and this
generates the sharp drop of $t_{relax}$. In other words, the band
implicitly ceases to exist and the central bank operates effectively in an infinitesimally narrow band. This provides new insight into target zone feasibility: if risk is too high, exchange
rate expectations are no longer anchored to the band and the
effectiveness of central bank intervention is greatly reduced. What the
central bank could do is therefore either (i) to reduce risk, which in
practice is often infeasible, or (ii) to increase the size of the target zone
which itself is bounded by the free-float exchange rate
volatility. The new size of the band would have to be large enough for
this new target zone to be ``heard''.

%

We can therefore also connect the threshold $\beta_e$ at which the
regime shift occurs to complete factor market integration: for lower
levels of $\beta$, the home fundamental exhibits an idiosyncratic
component anchored to its original dynamics that is stronger than its
tendency to converge to the target fundamental. Once this component is
overcome, the target zone ceases to exist and the currency starts
floating. This may also help explain why countries with a high
level of capital integration with the target currency may have higher
costs in maintaining a target zone. One implication of the suddenness
of the regime shift is that the relationship between capital
integration and the duration of the target zone is non-monotonic. This
is precisely what \citet{lera2015currency} illustrate with the case of
the Swiss Franc floor between 2011-2015.






The last contribution of our model lies in its generality, and how how it can replicate all the exchange rate densities presented by the established target zone literature.
We simulate central bank intervention by means of a symmetrized Euler scheme for stochastic differential equations. Since the original problem is a one-dimensional Neumann problem on the boundary $\partial D  = [-\bar{f}, \bar{f}]$, the regulated SDE can be written as: 

$$
f_t = \beta \int_0^t  \tanh (\beta f_s) ds + \sigma \int_0^t dW_s + \int_0^t\gamma(f_s) ds, 
$$
where the hyperbolic tangent is the nonlinear drift stemming from external risk\footnote{This is an equivalent representation of \eqref{dmps}: see \cite{arcand2018increasing} for further details.} and $\gamma(.)$ is the oblique reflection of the process on the boundary $\partial D$. This is the equivalent of the interventions, and we assume that for the unit vector field $\gamma$ there exists a constant $c$ so that $\gamma(x) \cdot \vec{n}(x) \geq c$ for all points $x$ on the boundary $D$. This can be interpreted as assuming bounded interventions. We use a regular mesh $[0,T]$ for the numerical simulation, for which the weak error is of order 0.5 when the reflection is normal (i.e. $\gamma = \vec{n}$), which is our case. We choose this method in order to obtain consistent Monte Carlo simulation of the resulting densities. The algorithm starts with $f_0 = 0$ and for any time $t_i$ for which $f_{t_i} \in D$ we have for $t  \in \Delta t =  t_{i+1}- t_i$ that: 
$$
F_t^{N,i} = f_{t_i}^N  + \hat{b}( f_{t_i}^N  )(t - t_i) +\sigma (W_t - W_{t_i}),
$$
as in the standard Euler-Maruyama scheme, and the nonlinear drift is approximated with a second-order stochastic Runge-Kutta method. If $F_{t+1}^{N,i}  \notin \partial D$, then we set: 

$$
f_{t+1}^{N}  = \pi^\gamma_{\partial D} (F_{t+1}^{N,i}  ) - \gamma(F_{t+1}^{N,i} ),
$$
where $\pi_{\partial D}(x) $ is the projection of $x$ on the boundary $\partial D$ parallel to the intervention $\gamma$. If $F_{t+1}^{N,i}  \in \partial D$, then obviously $f_{t+1}^{N} = F_{t+1}^{N,i}$.  For more references, see \cite{bossy2004symmetrized}. The exchange rate path is then obtained by setting $X^{N}_t = X^*(f_{t}^{N},T-t)$ for every $t \in [0,T]$. It is of fundamental importance to set $\Delta t$ equal to the update ratio given by $1/\alpha$ in our model, so that the increment of the simulated exchange rate path has the same updating time frequency as the central bank. \\

We can now discuss two kinds of interventions: the kind that intervenes by reflecting the process so that  it just stays within the band (sometimes called ``leaning against the wind''), and the pure reflection variety, which projects the fundamental process by an amount equivalent to how much the process would have surpassed the boundary. This distinction can also be understood as the amount of reserves the central bank has at its disposal in order to stabilize the fundamental process: the greater this quantity, the more likely it is that the intervention will be of the pure reflection type. We also assume that an intervention is effective instantaneously. This distinction also has important implications in the resulting exchange rate density: as shown in Figure \ref{dtbands}, given our characterization of risk, the greater the $\beta,$  the earlier  the central bank will have to intervene, given the fundamental's increased tendency to escape towards the bands. \\

\begin{figure}[htbp]
  \caption{Exchange rate densities, $\beta=0$ }
  \begin{tabular*}{\textwidth}{@{} c @{\extracolsep{\fill}} c @{}}
    \includegraphics[width=6cm]{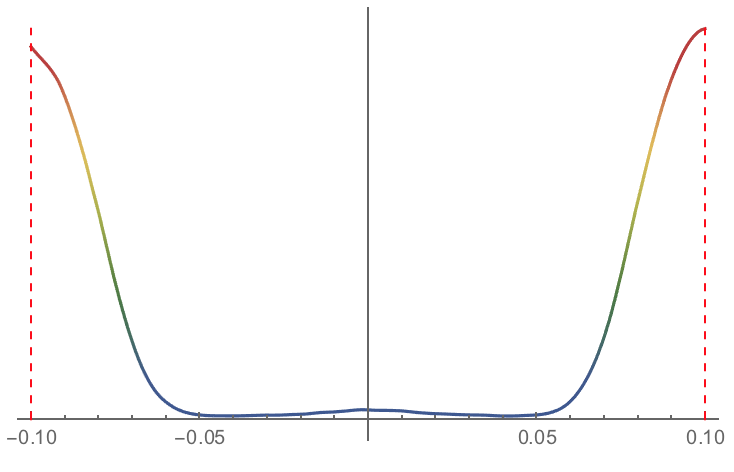} & \includegraphics[width=6cm]{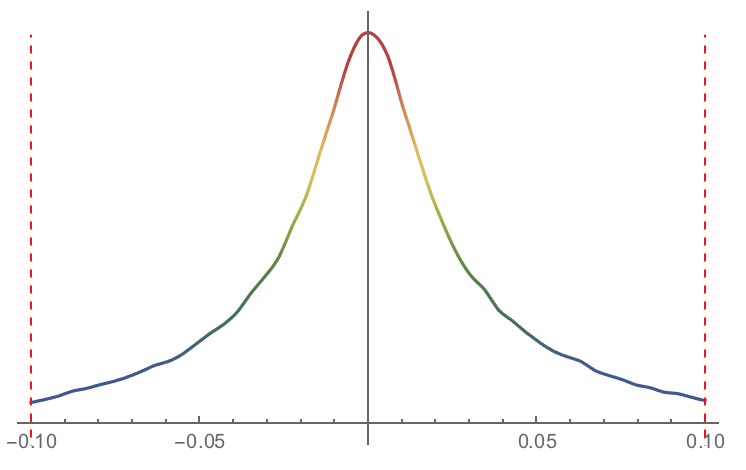} \\
    (a)  Marginal intervention, LAW & (b) Intramarginal intervention, pure reflection \\
  \end{tabular*}
  \label{dens1}
\end{figure}

\begin{figure}[htbp]
  \caption{Exchange rate densities, $\beta>0, \beta < \beta^e$}
  \begin{tabular*}{\textwidth}{@{} c @{\extracolsep{\fill}} c @{}}
    \includegraphics[width=6cm]{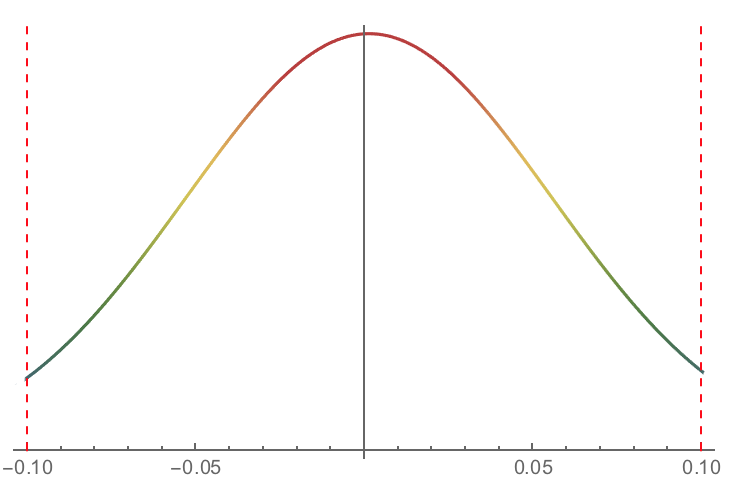} & \includegraphics[width=6cm]{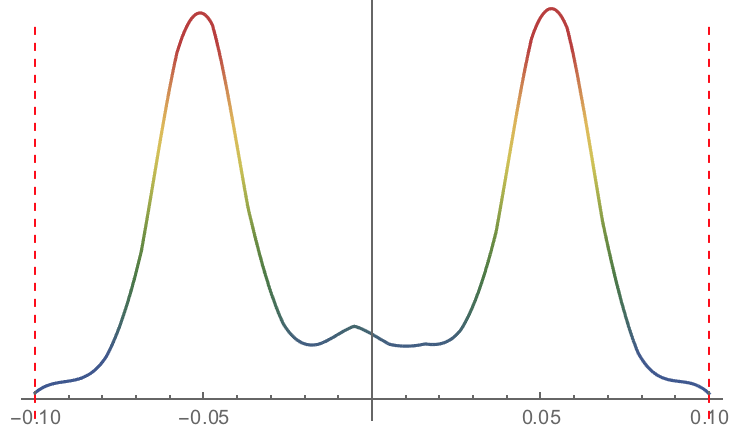} \\
    (a) Intramarginal intervention, LAW & (b) Two-regime intervention, pure reflection\
  \end{tabular*}
  \label{dens2}
\end{figure}

\begin{figure}[htbp]
   \caption{Exchange rate densities, $\beta>\beta^e$, }
   \begin{center}
    \includegraphics[scale=0.7]{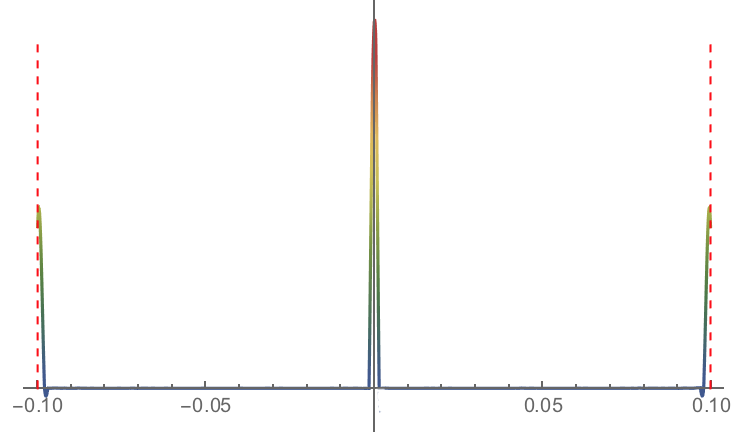} \\
Target band too narrow given the level of underlying risk.
  \end{center}
  \label{dens3}
\end{figure}

We present five possible scenarios by estimating Monte Carlo densities of the simulated exchange rate process: the first two correspond to the Gaussian case, where $\beta=0$ with each of the two intervention strategies. The densities are obtained by Monte Carlo simulation of $N$ sample paths, binning the data and limiting the bin size to zero to obtain the convolution density, then averaging over the $N$ realizations and interpolating the resulting points. For more references on the method, see \cite{asmussen2007stochastic}. For all figures $N$ is set to $5000$, $\sigma = 0.1, r= 0.5, \alpha = 200, T=3$ and the exchange rate target band to $\pm 10 \%$. We obtain a realization path for each of the two and obtain both U-shaped (corresponding to the standard Krugman model) and hump-shaped densities, corresponding to the \cite{dumas1992target} framework. The realized densities are plotted in Figure \ref{dens1}.  We then simulate the case in which $\beta >0$ but is not large enough to trigger the regime shift, each one with a different intervention strategy: in the marginal intervention case we obtain the two-regime density ($\beta = 5$), as in the \cite{bessec2003mean} framework, and in the intramarginal one we obtain a hump-shaped distribution as for all intramarginal intervention frameworks. These results are shown in Figure \ref{dens2}. 
The tendency $\beta$ of the fundamental to hit the boundary generates the two-regime shape, since even in a marginal framework the central bank will already intervene when at a distance from the bands: this is the case in which \emph{de facto} bands start to appear. Finally, we present the case in which $\beta$ is large enough ($\beta = 50$) to trigger the regime shift, and the band in fact ceases to exist: the tendency to escape leads to the fundamental process  constantly surpassing the boundary, honeymoon effects are impossible and pure reflection intervention concentrates most of the realizations around the initial level. The target zone is untenable and the central bank must either increase the bandwidth or drop the target zone altogether. As $N\to \infty$, the exchange rate density becomes a Dirac delta function around the initial value of the fundamental, displayed in Figure \ref{dens3}.




\section{Policy implications}
\label{s:policy}

A key contribution of this paper is to incorporate the exposure to external risk and associated destabilisation of exchange rate fundamentals in a target zone arrangement. 
Furthermore, we characterize how a target zone can be deemed feasible in order for central banks to enjoy honeymoon effects, which reduce the cost of intervention for achieving the set parity. The first policy implication of our framework lies in the fact that we are able to characterize the \emph{feasible band} for the target zone: from \eqref{eigelast} and the subsequent discussion one can obtain the \emph{minimum} size $| \overline{f} |  > 1/\beta$, which is directly related to the risk threshold that makes the target zone untenable. A central bank entering a target zone mechanism with a terminal exit time to another currency wants to limit the volatility of its exchange rate $X_t$ versus the anchor currency below the free-float level of the anchor currency. This provides us with a natural condition to the \emph{maximum} size of the band that the central bank can set as $|\overline{f}| \leq \sigma_z$,
where $\sigma_z$ is the long-term variance of the anchor currency. Considering the case of a target zone mechanism like ERM-II which has a band size of $\pm$15\%, the above condition implies that it is unlikely that the ERM-II bands will be breached. The Euro has an annualised volatility versus most currencies that is lower than the ERM-II target band. If there was \emph{no exposure} to external risk, it would be pointless to maintain a target zone with bands larger than $\sigma_z$. Our framework, therefore, gives central banks willing to enter a target zone arrangement clear guidelines on how wide the band should be.

The second policy contribution lies in establishing the concept of characteristic relaxation time $t_{relax}$, directly connected to the spectral gap \eqref{approx_bounds}. The relaxation time determines the minimum time a target zone must be maintained in order to ``feel'' the first effects of the home country central bank's actions.
The implications of this quantity for central bank policies are substantial: choosing an exit time below $t_{relax}$ would imply setting up an unfeasible target zone, as $ t_{relax}$ controls the minimum time necessary for agents to update their previously held exchange rate expectations, generating self-fulfilling expectations that create the honeymoon effect.
A further implication of this mechanism is that a central bank cannot adopt a target currency overnight with an arbitrary parity being the close of day value of the target exchange rate. 
In such a case, agents would not have had time to update their expectations and this would force the central bank to use a larger proportion of its assets (in the target currency) defending the parity level. This opens up many different avenues of enquiry into the expectation generation process of agents in foreign exchange markets. 
If $t_{relax}$ is the minimum time for agents to update their previously held exchange rate expectations, this means that greater shifts in higher degree of agent risk-aversion (higher $\beta$) will increase $t_{relax}$, which is an implication of our model. As shown by \cite{osler1995exchange} and \cite{lin2008}, this effect would work through the feasibility of the target zone in time
shifting speculators' horizons towards short term speculation, where $t_{speculation} \leq t_{relax}$. This is a natural outcome of honeymoon effects, which make intervention cheaper for central banks and harder for speculators \emph{after} $t_{relax}$. We find that $t_{relax}$ is increasing with the magnitude of the risk aversion shifts, for $\beta \leq \beta_e$. 
Our model does not deal with optimal choices: indeed, the only choice variable potentially available to the authorities is the time horizon $T$ by which the required parity needs to obtain. 
If one chooses an exit time which is lower than the required minimum time at which parity can be reached (the relaxation time), the target zone exit time is not feasible. However, setting a $T$ which is too high exposes one to increased business cycle risks, the dampening of which were a likely reason for entering a target zone in the first place. \\

Important policy relevance for our framework has emerged in a recent development for the Economic Community of West African States (ECOWAS) as well as for new entrants to ERM-II target zone like Croatia and Bulgaria. Both these target zones have seen their target zone exit times for participating countries lengthened in light of the COVID-19 pandemic. ECOWAS is planning to replace the current West African CFA Franc with a common currency, named Eco. The goal is for the 15 states to transition to the Eco via a target zone mechanism, similar to the ERM-II. One of the main concerns with the Eco was the short one year (and common) time horizon proposed for the target zone mechanism during discussions in 2019. This did not factor in the inherent idiosyncratic risk faced by individual West African central banks and ignored external risk factors. 
This translates directly to our framework, where the risk $\beta$ may generate a relaxation time $t_{relax}$ for individual states that may be larger than the proposed convergence time $T$. In light of the COVID-19 pandemic and its heterogeneous impact on the ECOWAS member states, the final time for convergence has been moved from 2021 to 2027, thus allowing a more realistic time frame for the ECO adoption. Our framework 
can guide the participating central banks away from unrealistic time horizons with potentially devastating consequences for their credibility, and for the overall process of creating the new common currency.  


Our framework has substantial implications for general managed exchange rate arrangements as well because it sheds light on the nature of external risk, and points towards different strategies that central banks can adopt in order to limit negative spillovers.
Our measure of fundamental risk can be used to fit a broad spectrum of global shocks which can destabilize exchange rate fundamentals. This naturally implies that central banks have to account for spillovers from the global financial cycle while managing their exchange rate arrangement. Negative spillovers that lead to capital flow reversals may induce exchange rate revaluations that are not in line with fundamentals, leading to increases in risk in \eqref{dmps2} that could challenge the feasibility of the target zone. The optimal policy response in such a situation is to increase the risk-sharing between the home currency central and the target currency central bank, which would reduce the pressure on the fundamental and bring the fundamental risk $\beta$ at a safe distance from the threshold $\beta_e$. This may take the form of swap lines or joint intervention in times of large exchange market pressure. Should such cooperation not be possible, home central banks may have to use macro-prudential and capital flow measures limiting capital flow volatility from translating to exchange rate volatility. In the case of target zones like ERM-II where there is explicit monitoring of the feasibility of the target zone by the target currency central bank, the same general principles guide the optimal policy response. The first-best outcome is to increase risk-sharing between home and target currency central banks. If this additional risk-sharing is not possible due to political considerations, technical support in surveillance of external risks and guidance in using macro-prudential tools is the next best outcome. \citet{dorrucci2020european} show that the European Central Bank (ECB) has made progress towards the second-best policy outcome by mandating that countries joining the Euro through the ERM-II have to join the European banking union and be subject to enhanced banking sector supervision.\footnote{The majority of the capital flow volatility realised by ERM-II countries in the 2000's was via banking sector capital flows rather than portfolio flows.} The use of swap-lines by the ECB for ERM-II countries during the COVID-19 pandemic is an example of pro-active policy ensuring the feasibility of these target zones. While provision of emergency swap lines is not standard practice, given legitimate concerns of moral hazard for new member states, our model shows how it may be useful in allowing for a successful convergence.\footnote{Bulgaria and Croatia officially entered ERM-II to replace their national currencies with the Euro in July 2020. The minimum convergence time $T$ to exit to the Euro is set at two years. Both the Lev and the Kuna have successfully pegged their currencies to the Euro over the last decade and may be considered as low $\beta$ countries at the time of their entry into ERM-II. However, in with \citet{fornaro2020finint}'s predictions, the accession of these countries to the Euro will be followed by an upgrade in the country ratings for foreign currency debt. These ratings upgrade at the time of the COVID-19 pandemic generates a higher probability of a capital flow surge and consequently a higher probability of a sudden stop into these countries. While capital flows might help with financing additional debt during the pandemic for these countries, it also generates a risk of these countries not meeting their fiscal criteria as well as destabilising the inflation expectations convergence process. At the time this paper is being written, it has been announced that Croatia will join the Euro in January 2023. 
}  \\

Lastly, 
in Appendix \ref{alt} we show how inflation expectations (key fundamentals in the determination of exchange
rates) of ERM-II countries follow a highly non-Gaussian distribution, compatible with our fundamental dynamics \eqref{dmps2}, and we provide alternative reduced form interpretations of $\lambda$ as a source of destabilizing risk. Furthermore, in this paper we choose to focus on a fundamental process that remains stationary in distribution around its long-run level, here normalized to 0 without loss of generality. This is the case for most target zone cases. However, if the fundamental was substantially misaligned from its long-run level, then the choice of a mean-reverting process could be more appropriate. The analysis of this case is
presented in Appendices \ref{ou} and \ref{soft}, where we fully solve both Ornstein-Uhlenbeck (O-U) and non-Gaussian, softly attractive dynamics. 
The latter can be of interest for researchers as an
alternative to the O-U process, since it allows one to again escape Gaussianity
and to model an ergodic process with light attraction towards its
long-run level, whilst maintaining analytical tractability.\\

\section{Conclusions}
\label{s:conclude}

Can one hear the shape of a target zone? The answer depends on whether the target zone is set up in a way which is feasible given the underlying fundamental risk that counteracts central bank efforts. In this paper we have explored the implications of extending exchange rate target zone modeling to non-stationary dynamics and heavy, non-Gaussian tails stemming from time-varying investor risk aversion, which lead to mean-preserving risk increases in the fundamental distribution. Our framework leads to a natural interpretation of target zone feasibility, driven by the interplay between two contrasting forces: a destabilizing effect driven by risk which pushes the exchange rate towards the bands, and a stabilizing diffusive force. For a given band, there is a maximum level of risk that allows one to ``hear'' the target zone.
We show how our model effectively endogenizes the presence of the bands by the exchange rate expectations, and how the interplay between risk and target band has key implications in the credibility of the zone itself, as well as the possibility of honeymoon effects. Intervention is shown to be both marginal and intramarginal, depending on how much the central bank ``hears'' the distance to the target zone band. The potential emergence of regime shifts, furthermore, can further erode the target zone credibility. This allows the methods employed in this paper to be applied to a wide range of situations.  
An important future extension of our work would be its empirical counterpart, consisting in the structural estimation of the model parameters and an explicit computation of the relaxation time, thus effectively providing the feasible band size as well as a lower bound for the necessary time for a central bank to reach the desired parity for its currency.

\newpage
\bibliographystyle{chicago} 
\bibliography{NLinear_ER}

\newpage
\begin{appendices}

\section{Monetary model of exchange rate determination}
\label{monetary}

Let us start with a standard flexible-price monetary model of exchange rate as in \citet{ajevskis2011target}. The money demand function is given as

\begin{equation}
m_t -p_t  = \theta^y y_t - \theta i_t+ \epsilon   \label{mon1}
\end{equation}
where $m$ is log of the domestic money supply, $p$ is log of the domestic price level, $y$ is the log of domestic output and $i$ is the nominal interest rate. $\theta_y$ is the semi-elasticity of the money demand with respect to output whereas $\theta_i$ is the absolute value of the semi-elasticity of money demand with respect to the domestic nominal interest rate and $\epsilon$ is a money demand shock. The second block is given by the expression for the real exchange rate $q$ which is defined as

$$
q_t = X_t + p^*_t -p_t    
$$
where $p^*$ is the log of the foreign price level. The third block of this model is the uncovered interest rate parity condition which in a linearised form is given by 

\begin{equation}    
 \mathbb{E}d X_t = (i_t- i^*_t) - \eta_t
 \label{eq:uip1}  
\end{equation}
where $\mathbb{E}d X_t$ is the is expectation of the exchange rate conditional on information available until time t and $i^*_t$ is the foreign interest rate. 
$\eta_t$ is a time-varying risk premium and is a consequence of risk-averse foreign investors who demand a higher compensation for holding home bonds and depends on investors’ risk aversion. 
Let us consider that investors face a standard problem of consumption of two bonds, home and foreign, with concave utility $U(c_t)$ discounted at $\gamma$. $B^h_t$ is the holding of home (small open economy) bonds $B^f_t$ is the holding of foreign bonds by a representative agent. Consumption and bond holdings in period $t$ and $t+1$ are given by the problem
  
\begin{eqnarray*}
\max_{c_{t+1},B^h_t, B^f_t} & &  \sum_{t=0}^\infty \gamma^t U(c_t) \\
    c_t &=& B^h_t +X_t B^f_t \\
    \mathbb{E}[c_{t+1}] &=& (1+i_t) B^h_t + \mathbb{E}[X_{t+1}](1+i^*_t) B^f_{t}.
\end{eqnarray*}
Solving this problem leads straightforwardly to Eq. \eqref{uipnew}. Using equations \eqref{mon1}, \eqref{eq:uip1} and \eqref{uipnew}, we recover the monetary model of the exchange rate as given by
 
\begin{eqnarray*}
\theta\mathbb{E}d X_t  \underbrace{\underbrace{- \theta^y y_t +q_t+p^*_t + \theta i^*_t-\epsilon}_{v_t} + \theta \eta_t + m_t}_{f_t} = X_t&&3
\end{eqnarray*}
\begin{eqnarray}
    X_t & =&  \theta \mathbb{E}_t \{ dX_t \}+ v_t + \theta \eta_t+  m_t \nonumber \\
   & = & \theta \left ( \mathbb{E}_t \{ dX_t \} + \frac{d\mathbb{Q}}{d\mathbb{\mathbb{\tilde{Q}}}} \right )  + v_t + m_t. \label{model1}
\end{eqnarray}
In the literature the velocity $v_t$ is usually modeled as a Brownian motion. In order to include $\eta_t$, the  time-varying risk premium, we modify the Brownian motion to include the incremental risk generated by varying investor risk aversion by means of \eqref{pert}. Intervention happens at the boundaries $\underline{f}, \overline{f}$, at which the central bank undergoes infinitesimal adjustments of money supply $m_t$  in order to keep the fundamental in the band. Equations \eqref{dmps}, \eqref{EXCHA} and the boundary conditions then follow directly.


\section{Proof of Proposition 1}\label{derivations}

Using It\^o calculus, Eq.(\ref{EXCHA}) can be written as follows:

\begin{equation}
\label{EX11}
  \partial_t X(t,f) + {\sigma^2 \over 2} \partial_{ff} X(t,f) + \beta \mathcal{B}  \partial_f X(t, f)  - \alpha X(t,f) = - \alpha  f.
\end{equation}
Note the presence of the additional term $\partial_t X(f) $ in
Eq.(\ref{EX11}) which does not appear when one focuses only on
stationary situations.  As shown in \cite{arcand2018increasing}, Eq.(\ref{EX11}) can be written equivalently as the nonlinear partial differential equation given by
  
  \begin{equation}
  \partial_t X(t,f) + {1 \over 2} \partial_{ff} X(t,f) + \beta \tanh(\beta f)   \partial_f X(t, f)  - \alpha X(t,f) = - \alpha  f.   \label{EX111}
\end{equation}
Using the superposition principle, the solution of \eqref{EX111} can be written as the sum of the time-independent stationary solution and the non-stationary solution:

\begin{equation}
  X(\tau, f) = X^{*}(\tau ,f) + X_S(f). \label{totsol}
\end{equation}
For the derivation of the stationary solution, we first introduce
the following \emph{Ansatz}:
  \begin{equation}
    \label{DARBOUX}
    \partial_{f} X(t,f) = Y(t,f)/\cosh(\beta f).
  \end{equation}
This leads to the following transformations:
\begin{eqnarray*}
X &=& \frac{Y}{\cosh(\beta f)}, \\
\beta \tanh(\beta f)  \partial_f X &=& \left[ \beta \frac{\sinh(\beta f)}{\cosh(\beta f) } \right] \left[- \beta \frac{\sinh(\beta f) }{\cosh^{2}(\beta f) }  + \frac{\partial_f Y}{\cosh(\beta f) } \right]  \\
&=&   -\beta^{2} \frac{\sinh^{2} (\beta f) }{\cosh^{3}(\beta f)} Y  + \beta  \partial_f Y \frac{\sinh(\beta f) } {\cosh^{2}(\beta f)} ,\\
\frac{1}{2}  \partial_{ff} X &=& \frac{1}{2\cosh(\beta f)} \partial_{ff}Y - \frac{\beta  \sinh(\beta f) }{ \cosh^{2}(\beta f)} \partial_f Y- \frac{ \beta^{2}}{2}   
\frac{Y}{\cosh(\beta f)}+  \beta^{2} \frac{\sinh^{2}(\beta f) }{\cosh^{3}(\beta f)}Y,
\end{eqnarray*}

which substituted in \eqref{EX111} yield the following linearization:
  \begin{equation}
    \label{EX2}
    \partial_t Y(t,f)+  {1\over 2} \partial_{ff} Y(t,f) - \left[ {\beta^{2} \over 2}  + \alpha \right]Y(t,f)= - \alpha f \, \cosh (\beta f).
  \end{equation}
  Setting $\partial_t  = 0$ one obtains a nonlinear ODE in $f$ which has the closed form solution as given by
  \begin{equation}
  \label{YPH}
  \left\{
    \begin{array}{l}
      {\cal Y}_1(f) = \exp\left\{ +\sqrt{ \left[ \beta^{2}   +  {2\alpha   }\right]} f \right\} ,
      \\ \\
      {\cal Y}_2 (f)= \exp \left\{-\sqrt{ \left[ \beta^{2} +  {2\alpha   }\right]}f \right\} , \\ \\
      Y_P(f)= \frac{2 \alpha  \left(f \left(2 \alpha\right) \cosh (\beta f) +2
    \beta  \sinh (\beta f )\right)}{\left(2 \alpha\right)^2}
  \end{array}
  \right.
\end{equation}
which is the sum of the general solution (two opposite-sided exponentials) and a particular solution. Obtaining the general solution is a simple exercise and thus omitted, while the particular solution requires a little more attention. We introduce another {\it Ansatz}:

$$
\left\{
\begin{array}{l}
Y = \left[R f \cosh(\beta f) + S\sinh(\beta f) \right] , \\ \\ 
\partial_f Y = \left[ R \cosh(\beta f) + Rf\beta \sinh(\beta f)  + S \beta \cosh(\beta f) \right],\\ \\ 

\partial_{ff} Y = \left[ 2 R\beta  \sinh (\beta f) + Rf\beta^{2} \cosh (\beta f)  + S \beta^{2}  \sinh(\beta f) \right.
\end{array}
\right.
$$

\noindent  We therefore have:
$$
\begin{array}{l}
 \frac{1}{2} \partial_{ff} Y - \left[ \alpha + \frac{1}{2} \beta^{2} \right] Y + \alpha f \cosh(\beta f) = \\ \\ 
 \left[\frac{1}{2} R \beta^{2}  -  \left[ \alpha + \frac{1}{2}\beta^{2} \right] R+\alpha \right] f \cosh(\beta f) + 
\left[ R \beta + \frac{1}{2} S \beta^{2} - \left[ \alpha + \frac{1}{2} \beta^{2} \right]  S\right] \sinh(\beta f) =0.
\end{array} $$

\noindent Matching coefficients we obtain:

$$
\left\{
\begin{array}{l}
 \left[\frac{1}{2} R \beta^{2} -( \alpha  + \frac{1}{2} \beta^{2} )R+ \alpha \right] =0, \\ \\ 
\left[ R \beta + \frac{1}{2} S \beta^{2} -\left[  \alpha + \frac{1}{2} \beta^{2} \right]S\right] =0 .
\end{array}
\right.
$$

\noindent which implies $ R =  1, S = \frac{\beta }{ \alpha}$ and thus

$$
Y_P = \frac{\alpha f \cosh(\beta f) + \beta \sinh (\beta f) }{\alpha} .
$$
\noindent Inverting the transformation back to $X$ one obtains \eqref{XPH}. 
 \noindent In Eqs. (\ref{XPH}) athe pair of constants
$A$ and $B$ can be determined by smooth fitting at the bounds $\underline{f}=-\overline{f} $: 

$$
  \partial_f X_S(f) \mid_{f = \underline{f}} \,\, =\,\;  \partial_f X_S(f) \mid_{f= \overline{f} }\,\,   =\,\; 0.
$$
The two constants of integration $A$ and $B$ can be obtained
in closed form but their expression is lengthy and is therefore
omitted. \\

We now turn to the non-stationary dynamics. At a given time horizon $t=T$,
we fix the predetermined non-stationary part of the exchange rate at exit time $X(T,f)=0$. In terms of the
backward time $\tau = T-t$, we write the transformation
$ X^{*}(\tau, f) = Y^{*}(\tau, f) / \cosh (\beta f) $. We then need to solve the following nonlinear boundary value problem:

  \begin{equation}
    \label{EX11}
    \left\{
      \begin{array}{l}
        \partial_{\tau} X(\tau, f) - {1\over 2} \partial_{ff} X(\tau, f) - \beta \tanh(\beta f)  \partial_f X(\tau,  f)  + \alpha X(\tau ,f) = + \alpha f, \\
        X(0, f) =0 \\
        \partial_f X^{*}(\tau ,f ) \mid_{f= \underline{f} } =0  \\
         \partial_f X^{*}(\tau ,f ) \mid_{f= \overline{f} } =0 
      \end{array}
    \right.
  \end{equation}

  \noindent Writing $X(\tau, f) = X^{*}(\tau ,f) + X_S(f)$,
  Eq.(\ref{EX11}) implies:

  \begin{eqnarray}
    \label{EX22}
     & &  - {1 \over 2} \partial_{ff} X_S(f)- \beta \tanh(\beta f)  \partial_f X_S(f)  + \alpha X_S(f) = \alpha  f, \\
     & &    \partial_{\tau}X^{*}(\tau ,f) - {1 \over 2} \partial_{ff} X(\tau ,f)-\beta \tanh(\beta f)  \partial_f X(\tau, (f)  + \alpha X^{*}(\tau, f) =  0 \nonumber 
  \end{eqnarray}

  \noindent While the first line in Eq.(\ref{EX22}) has already being
  solved, the second line needs now to be
  discussed. Writing again
  $ X^{*}(\tau, f) \cosh (\beta f) := Y^{*}(\tau, f)$, we obtain:

  \begin{equation}
    \label{EX33}
    \partial_{\tau}Y^{*}(\tau, f) - \frac{1}{2} \partial_{ff} Y^{*}(\tau, f)   +  \left[  {\beta^{2} \over 2} + \alpha \right]  Y^{*}(\tau, f) =0.
  \end{equation}

  \noindent The boundary conditions given by Eq.(\ref{SM})
  impose:

  \begin{equation}
    \label{SM1}
    \left\{
      \begin{array}{l}
        \partial_f X^{*}(\tau ,f ) \mid_{f= \underline{f} } =0  \quad \Rightarrow \quad \left\{ \left[ \partial_f Y^{*}(\tau ,f ) \right] - \beta \tanh (\beta f) Y^{*}(\tau ,f )\right\} \mid_{f= \underline{f} } =0,\\ \\
        \partial_f X^{*}(\tau ,f ) \mid_{f= \overline{f} } =0  \quad \Rightarrow \quad   \left\{  \left[ \partial_f Y^{*}(\tau ,f )\right] - \beta \tanh (\beta f) Y^{*}(\tau ,f )\right\} \mid_{f= \overline{f} } =0.
      \end{array}
    \right.
  \end{equation}


\noindent We express the solution $Y(\tau, f) $ as
$Y^*(\tau, f) = \phi(\tau) \psi(f)$, and proceed to solve this
equation by separation of variables and expansion over the basis of a
complete set of orthogonal eigenfunctions. 
  \noindent We solve \eqref{EX33} by separation of variables and
  expansion over the basis of a complete set of orthogonal
  eigenfunctions. The solution can be expressed as
  $Y^*(\tau, f) = \phi(\tau) \psi(f)$, and therefore we can write it
  as

$$
\frac{\dot{\phi}(\tau)}{\phi(\tau)} = \lambda_k =
\frac{1}{2}\frac{\psi''(f)}{\psi (f)} - \rho
$$
where $\rho = \left[  {\beta^{2} \over 2} + \alpha \right] $. \\
\\
The time-dependent part solves to
$\psi(\tau) = \exp ( \tau \lambda_k)$, and the fundamental-dependent
part can be written as the ordinary differential equation

$$
\psi''(f) - 2(\lambda_k + \rho) \psi(f) = \psi''(f) + 2 \Omega_k^2
\psi(f) = 0.
$$
Solving for
$\psi$ one obtains the eigenfunctions

$$
\psi_k(f) = c_1 \cos \left (\sqrt{2} \Omega_k f_t \right ) + c_2 \sin\left (\sqrt{2}
\Omega_k f_t \right ).
$$
Sturm-Liouville theory allows us to state that on the interval
$[-\overline{f}, +\overline{f}]$, one has a complete set of orthogonal
eigenfunctions $\psi_{k}(f)$ satisfying Eq.(\ref{SM})
which form an orthogonal basis for the $2\bar{f}$-
well-behaving functions space. Smooth-fitting conditions impose
$c_1 =0, c_2=1$ and we obtain the form of the eigenfunctions

\begin{equation}
  \label{ORTHO}
  \psi_{k}(f) = \sin\left(\sqrt{2} \Omega_k  f \right) \in [ \underline{f}, \overline{f}], k=\mathbb{N}^+,
\end{equation}
where each eigenvalue $\Omega_k$ solves the transcendental equation:

\begin{equation}
\sqrt{2} \Omega_k \cot \left (\sqrt{2} \Omega_k  \overline{f} \right )
= \beta  \tanh( \beta \overline{f} ). 
    \end{equation}
as given
by \eqref{ORTHO}. By regularity of the Sturm-Liouville problem we know that the eigenvalues are real and span a discrete spectrum:

$$
\left\{ \Omega_k\right\}:= \left\{ \Omega_k (\beta , \overline{f})\right\},k\in \mathbb{N}^+.
$$  

\noindent and can therefore be ordered as:

$$
\Omega_1(\beta , \overline{f})< \Omega_2 (\beta , \overline{f})<
\cdots  \Omega_k (\beta , \overline{f}).
$$

The Fourier coefficients follow in their standard
form, using the stationary equation $X_S(f_t)$, and we finally obtain \eqref{EQTRANSIENT}.

\section{Proof of Proposition 2}
\label{honeymoon}

We now briefly discuss the connection between risk and the honeymoon
effect, and how such effects cannot be be obtained when the destabilizing effects of risk shocks in the fundamental are too strong. For illustrative purposes,
let us consider a baseline case of our model in a
symmetric band $[ - \overline{f}, \overline{f}]$ around the parity
$0$, and let us compare our model with the standard
Gaussian one. Omitting time dependency, we have again the framework
given by

$$
\label{KRUGB}
X = f+ \frac{1}{\alpha} {\mathbb{E} \left\{ dX\right\} \over dt},
$$
which leads to the following couple of PDEs, depending on the form of
the fundamental process.
$$
\begin{cases}
  X = f + {1\over 2} \partial_{ff}[ X(f) ]  \qquad \qquad   \qquad \qquad  \qquad \qquad (\text{Gaussian}), \\
  X = f + {1\over 2} \partial_{ff}[ X(f) ] + \beta \tanh (\beta f)
  \partial_f [ X(f)] \qquad ({\rm Ours}).
\end{cases}
$$
We now focus on the stationary regime for which get the general
solutions:
$$
\begin{cases}
  X(f) =f + A_0\sinh( \rho_0 f),  \qquad \qquad   \qquad \qquad  \qquad \qquad (\text{Gaussian}),\\
  X(f) = f + A _{\beta}{ \sinh( \rho_{\beta} f) \over \cosh (\beta
    f)}, \qquad \qquad \qquad \qquad \qquad \qquad \qquad ({\rm
    Ours}),
\end{cases}
$$
\noindent where $ \rho_{\beta} = \sqrt{\beta^{2} + 4 \alpha }$ and
$A_{\beta}$ is a yet undetermined amplitude. We now apply the smooth
fitting procedure at the target level $+\bar{f}$\footnote{Due to the
  symmetry, we have here only one amplitude$A$ to determine since only
  one boundary needs to be considered.}.

For the standard Gaussian framework we have $ X(f) \mapsto X_0(f) =  f + a \sinh (\rho_0 f) $, since $\beta =0$ and consequently $ \rho \mapsto \rho_0 :=   \sqrt{ \frac{2\alpha }{ \sigma^{2}} } $. We therefore have :

$$
X_0(f) = a \tanh(\rho_0  f) +  f, \qquad \qquad \rho_0 =   \sqrt{ \frac{2\alpha }{\sigma^{2}} } ,
$$

\noindent which is the same result as in the standard Gaussian models. In particular, denote $W \in \mathbb{R}$ the contact point with the target boundary $\pm \bar{f}$, we have

\begin{equation}
\label{K1}
\left\{
\begin{array}{l}
  \bar{f}= X_0(W) \quad \Rightarrow \quad  F= W +   a \tanh(\rho_0  W) , \\ \\ 
 
 0 = 1  + \rho_0 a \cosh(\rho_0 W)

 \end{array}
 \right.
\end{equation}
because of the smooth-fitting boundary conditions on the first derivative. From the second line in  Eq.(\ref{K1}), we conclude immediately that:

$$
a =  \frac{-1}{\rho_0 \cosh( \rho_0 W)}.
$$

\noindent and accordingly,  we end with:

\begin{equation}
\label{K2}
X_0(f) =  f -\frac{\sinh(\rho_0 f) }{\rho_0 \cosh(\rho_0 W) }  
\end{equation}

\noindent Furthermore, we can  verify  that $W\in \mathbb{R}^{+}$ for all  values of the parameter $\rho_0 >0$. Eq.(\ref{K2}) implies that:

$$
 W - F=   \frac{\tanh (\rho_0 W) }{\rho_0}.
$$

\noindent It can be immediately seen that the last equation always possesses a single solution $W \in \mathbb{R}^+$. Let us now examine the paper's main framework, the case where $\beta > 0$. In this case, for a target zone with band size $F$ and a smooth contact point $W$,    we have:

\begin{equation}
\label{DMPS1}
\left\{
\begin{array}{l}
 F = W + a \frac{\sinh (\rho W) } {\cosh (\beta W)} + \omega \tanh( \beta W)
\\ \\
0= 1 + \frac{ a }{\cosh(\beta W) }  \left[\rho  \cosh(\rho W)  - \beta \sinh(\rho W) \tanh(\beta W)\right] + \frac{ \omega \beta }{\cosh^{2}(\beta W)}.
\end{array}
\right.
\end{equation}

\noindent The second line of the last equation implies:

$$
a = - \frac{  \cosh^{2} (\beta W) +  \beta \omega}{ \cosh (\beta W) \cosh (\rho W)\underbrace{ \left[ \rho \tanh( \rho W)  - \beta \tanh(\beta W) \right]}_{:= \Delta } }.
$$

\noindent From the last line, let us consider the equation $\Delta =0$. First we remember  from the very  definition that $\rho \geq \beta$ and hence the equation:

$$
\quad \frac{\rho}{\beta} \tanh(\rho W) = \tan(\beta W)  \qquad \Leftrightarrow\qquad  \Delta = 0.
$$

\noindent Since $\frac{\rho}{\beta} >1$ the last equation necessarily admits a solution $\pm W_c$.  Note in  addition that for  a couple of $\beta$ such that $\beta _1 < \beta_2$,  we have:

\begin{equation}
\label{ORDER}
\beta _1  > \beta _2 \quad \Leftrightarrow \qquad W_{c,1} < W_{c,2} 
\end{equation}

\noindent and for $\beta \rightarrow \infty$, we have $W_c \rightarrow 0$. Now  from $W$ solving  the first line of Eq.(\ref{DMPS1}), we may have the alternatives:

\begin{equation}
\label{ALTERNAT}
\begin{array}{l}
a) \quad W_c < W ,
  \\ \\
b)\quad  W_c \geq W.
\end{array}
\end{equation}
Since the contact point $W_c$ decreases as $\beta$ increases, there exists a $\beta^e$ for which $W_c = W$. For all $\beta > \beta_e$, standard boundary fitting techniques cannot be applied as in the Gaussian case, and hence the limit $W= W_c$ explains the regime shift observed in the spectrum. This is due to the fact that for large $\beta$ the honeymoon effect range becomes effectively large enough to preclude the possible existence of a target zone. The eigenvalue jump can be obtained by examining the boundary conditions given by \eqref{SM1}, and separating the contribution to the smooth-fitting at the boundaries given by the eigenfunction from the one given by the fundamental drift and obtain:

$$
  \frac{\partial_f\sin \left (\sqrt{2} \Omega_k f_t \right)}{ \sin \left (\sqrt{2} \Omega_k f_t \right)} - \beta \tanh(\beta f_t) =0.
  $$
  By noticing that $\beta \tanh( \beta f)$ is equivalent to $\beta \mathcal{B}$, and is thus the mean-preserving spread caused by increases in fundamental risk, one obtains 
  $$
  \frac{\partial_f \psi_k (\Omega_k, f_t)}{\psi_k (\Omega_k, f_t)}  -  \text{MPS}(\beta, f_t) = 0 
  $$
and we obtain \eqref{eigelast}.

\section{Noise sources driving the fundamental}
\label{assumptions}
\noindent Let us now assume that the fundamental is driven by a couple
of noise sources, namely i) composite shocks $v_t$ and ii)
fluctuations in the money supply $m_t$, given by Gaussian noise around
a drift $\mu$. We therefore add another source of noise, but we are
not necessarily increasing the risk in the fundamental process. We
then have

\begin{equation}
  \label{FONDOS1}
  \left\{
    \begin{array}{l}
      df_t = \sigma_{1} dW_{1,t} + dm_t, \\ \\
      dm_{ t}= \mu dt + \sigma_{2} dW_{2,t}, \qquad m_{t=0} =m_0.
    \end{array}
  \right.
\end{equation}
where the noise sources $dW_{1,t}$ and $dW_{2,t}$ are two independent
White Gaussian Noise (WGN) processes.  We then obtain $f_t$ as a
Gaussian process, since trivially

$$
df_t =\mu dt + \sqrt{\sigma_1^2+\sigma_2^2} dW_t
$$
and we are exactly in the standard framework (in the literature
usually $\mu = 0$), only with a change in variance. If however we wish
to incorporate a general increase in risk, and one that may represent
the force that was discussed in Section 2, we can write the
following more general framework:

$$
  \label{FONDOS}
  \left\{
    \begin{array}{l}
      df_t = \sigma_{1} dW_{1,t} + dm_t, \\ \\
      dz_{\beta, t}= \zeta (\beta; z_t)dt + \sigma_{2} (\beta) dW_{2,t}, \qquad z_{t=0} =0.
    \end{array}
  \right.
$$

\noindent where $\beta\geq0$ is a control parameter and the repulsive
drift $\zeta(\beta; z) = - \zeta(\beta ; -z) <0$ models an extra risk
source via a dynamic zero mean process. We parametrize risk with
$\beta$, and therefore $\beta =0$ simply implies
$ \sigma_2 (\beta) = \zeta (0; z_t)=0$ implying that the process is
Gaussian and driven entirely by the composite shock process. Our
candidate for $\zeta$ is the DMPS process:

$$  \label{BETADMPS}
  \begin{array}{l}
    df_t=  \sigma_{1} dW_{1,t}   + dz_t =  \beta \tanh (\beta  z_t)   dt  + \sigma_1 dW_{1,t} +  \sigma_{2}(\beta) dW_{2,t}   \\\\

    \qquad  \qquad \qquad  \qquad \qquad  \qquad \qquad\Downarrow  \\ \\

    dz_t =  \beta  \tanh (\beta z_t) dt +  \left[  \sqrt{\sigma_{1}^{2} + \sigma_{2}^{2} (\beta) } \right]dW_{t}, \quad z_{t=0} =0.

  \end{array}
$$

\noindent where we used the fact that the difference between two
independent WGN's is again a WGN with variance as given in the
previous equation.  Alternatively one may formally write:

$$  
\label{ALTER}
\begin{array}{l}
  df_t = \sigma_{1} dW_{1,t}   +   \beta \tanh \left[ \beta \overbrace{( f_t - \sigma_1 W_{1,t})}^{z_t} \right]   dt  +  \sigma_{2}(\beta) dW_{2,t}  = \\ \\

  \qquad \qquad    \beta \tanh \left[ \beta \overbrace{( f_t - \sigma_1 W_{1,t})}^{z_t} \right]   dt  + \left[  \sqrt{\sigma_{1}^{2} + \sigma_{2}^{2} (\beta) } \right]dW_{t},
\end{array}
$$
\noindent Using the initial equation (\ref{EXCHA}) and the previous
equation and applying It\^o's lemma to the functional $X(f_t, t) $, we
obtain:
\begin{equation}
  \label{ITO}
  \begin{array}{l}
    {(1-r) \over  \alpha }\ \left\{  \partial_t X(f, t )  + \partial_f X(f, t ) \underbrace{  \mathbb{E} \left\{  \beta \tanh \left[ \beta ( f_t - \sigma_1 W_{1,t}\right]    \right\}}_{= \beta \tanh \left[ \beta ( f)\right]}   +\left[  \sigma_{1}^{2} + \sigma_{2}^{2} (\beta)  \right]\partial_{ff} X(f, t)\right\} =
    \\ \\
    \qquad \qquad \qquad X_t- r\, f_t \end{array}
\end{equation}

\noindent In the last Eq.(\ref{ITO}), the under-brace equality follows
since all odd moments in the expansion of the hyperbolic tangent
vanish and the
$\tanh(x)$ is itself an odd function. Now, normalizing as to have
$\left[ \sigma_{1}^{2} + \sigma_{2}^{2} (\beta) \right]
=\sigma^2 $, we are in the nominal setting of our paper.

\section{Alternative interpretations of risk} 
\label{alt}

\begin{figure}[htbp]
    \begin{center}
        \includegraphics[width = 8cm]{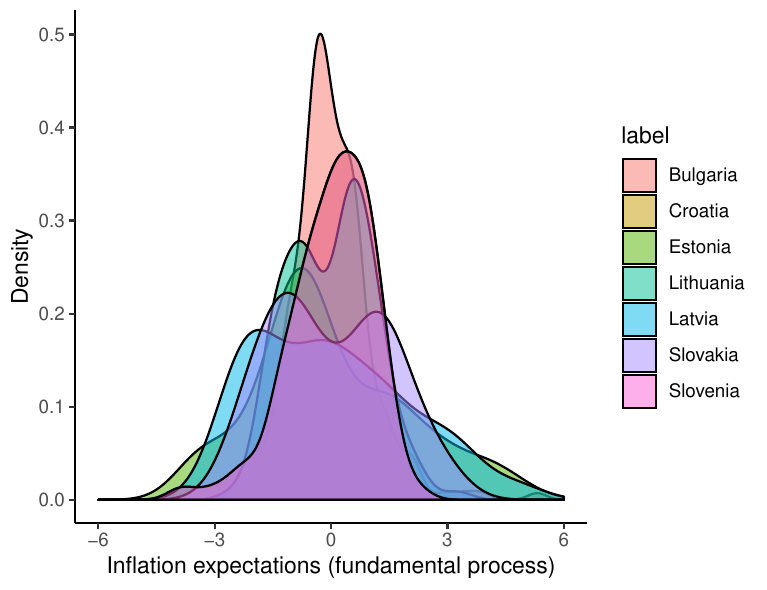}
         \includegraphics[width = 7cm]{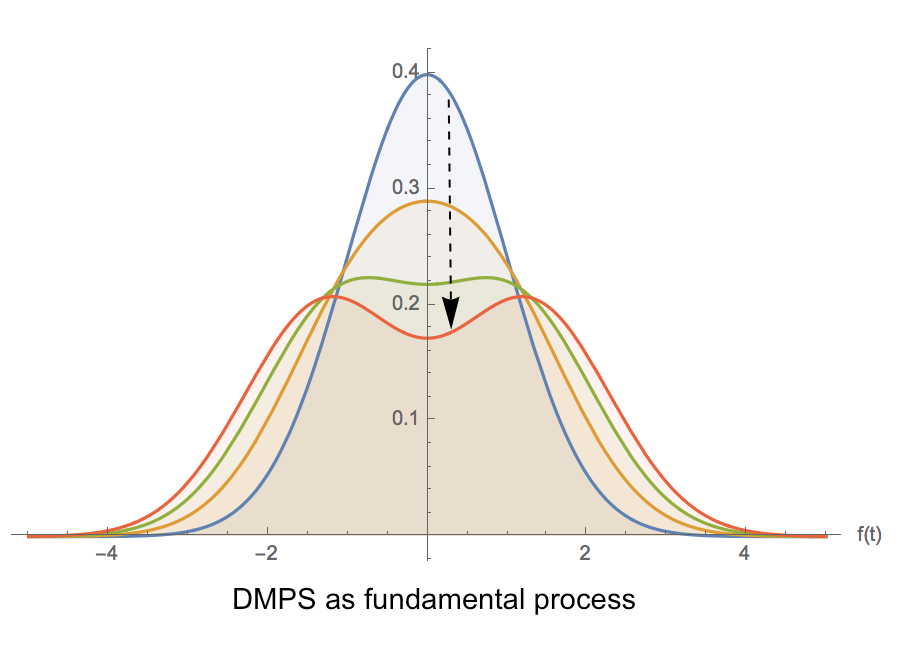}
    \caption{Estimated densities of the fundamental process (inflation expectations) for ERM-II currencies}
    \label{fig1}
    \end{center}    
    \floatintro{(Left panel) Centered difference between Euro area inflation expectations and target zone country inflation expectations, for the time each currency was in the target zone with the Euro. Kolmogorov-Smirnov tests greatly reject each hypothesis of Gaussianity. The data for inflation expectations comes from the Euro Commission's Joint Harmonised Consumer Survey. For more details we refer to \citet{arioli2017eu}. Bulgaria and Croatia have only recently acceded to joining the Euro and the data for them is backward-looking to give the reader a sense of pre-target zone differences in inflation expectations. (Right panel) Transition densities of the fundamental process with mean-preserving spreads at time $t=1$, each with risk increases in the direction of the arrow.}
\end{figure}

Destabilization is intrinsically connected to risk in the fundamental process. Besides the structural interpretation of risk as stemming from time-varying investor risk aversion, one could think of a variety of other interpretations for the parameter $\lambda$ of increasing risk, which generates mean-preserving spreads in the density of the exchange rate fundamentals. A quick glance at the left-hand panel of Figure \ref{fig1} shows that the difference in inflation expectations, one of the key fundamentals in the determination of exchange rate target zones, is undoubtedly non-Gaussian, exhibits substantially heavier tails and presents bimodal tendencies stemming from both inflationary and deflationary pressures shifting probability away from the center. Such risk dynamics cannot be represented by the variance of Gaussian fluctuations, as they cannot affect the distribution tails, but rather requires the presence of forces that increase the tendency of the fundamental process to escape its long-run level. The right panel of Figure \ref{fig1} shows the transition density of the DMPS process at an arbitrary time for increasing risk. The DMPS density with $\lambda$ parameter fit by maximum likelhood is a better fit for the empirical densities for each of the densities shown in Figure \ref{fig1}.\\

Another way of interpreting of the risk parameter of our framework could be via the presence of capital flows, especially in how the magnitude and the drivers of capital flows matter in determining the stabilisation effects. First, capital flows may be driven by push factors creating cycles of bonanzas and sudden stops seen with New Member States. \citet{hansson2013} argue that capital flow dynamics were the key driver for cyclical developments in the Baltic ERM economies. This is might be a issue for a small target zone country if the capital flows generate excess appreciation or depreciation pressure weakening the feasibility of the target zone. 
This is particularly problematic if there is a sudden stop with reallocation of capital flows to more productive economies in the target zone as seen during the Eurozone crisis \citep{ghosh2020currency}. Furthermore, assuming absence of macro-prudential tools, capital flow volatility may generate foreign exchange intervention volatility inside the target zone, as the use of interest rates as a monetary policy tool can generate further pro-cyclicality in capital flows. This nexus between capital flows and target zone management may destabilise the convergence in the inflation process of the target zone country. This is the key source of additional risk in our setting. Let's consider the real interest rate version of the UIP condition:

$$
\mathbb{E} \left \{d X_t \right \} = ( r_t  - r^*_t) dt + \mathbb{E} \left \{ d\pi_t - d\pi^*_t \right \},
$$
where $\pi^*$ is the target country's inflation measure and $\pi$ is home inflation. 
If there are high capital inflows that need to be counteracted by (unsterilised) intervention, this would generate a lower real interest rate of financing by putting downward pressure on $r_t$. This additional supply of credit is likely to increase the $\mathbb{E} \left \{ d \pi_t \right \}$ This would require an interest rate response by the national central bank, in the absence of macro-prudential tools. We can see that in this particular case, increasing interest rates may be pro-cyclical to capital flows as long as the inflation process responds positively to the interest rate hike, causing a loss of monetary autonomy if the process is self-reinforcing. A destabilizing outcome of this setting would be if the inflation process does not respond to the interest rate moves and causes an outflow of capital flows. This would jeopardise the feasibility of the target zone and could cause the gap between $r_t$ and $r^*_t$ to become larger than before entering the target zone. The standard approach of modeling risk in the target zone does not consider the risk stemming from the currency union itself. If the target currency union has real interest rate changes through lower expected inflation surprises, it will also affect the stability of target zone by the capital flow mechanism we have described.\footnote{For simplicity, we do not consider the currency union having positive inflation surprises, even though in a low real interest rate setting, it may lead to capital flows to the target zone currency. This mechanism can be amplified by presence of multiple currencies in the target zone with cross-currency constraints on movement versus the target currency \citep{serrat2000}. } Lastly, we note that our characterization of risk as destabilizations caused by capital flows can be further extended to any source of external risk, and our model framework would still apply. \\

\section{Attracting drift: mean-reverting dynamics}

\label{ou}
\begin{figure}[htbp]
  \caption{Mean-Reverting Stationary
    Dynamics}\label{fig:betas_spectrum}
  \includegraphics[scale=0.8]{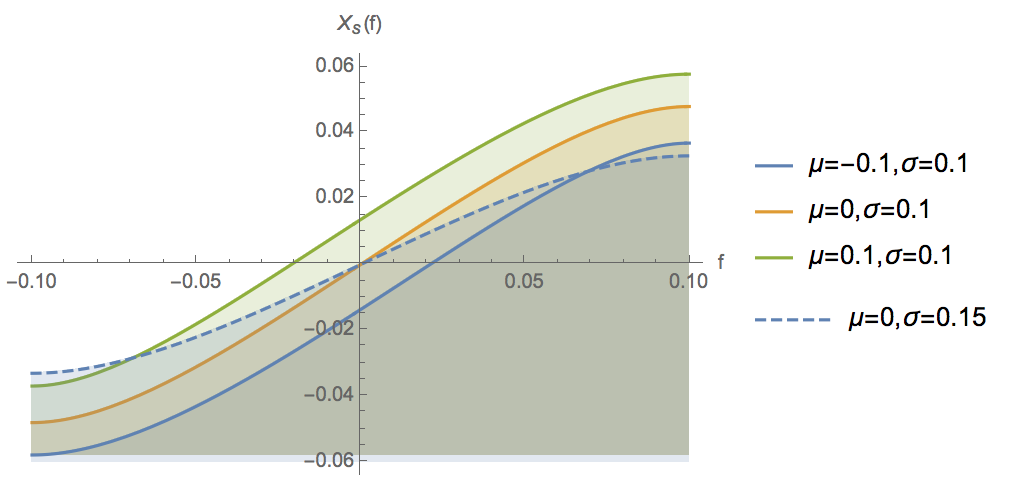}
  \label{oustat}
\end{figure}

A fully similar discussion can be done for mean-reverting fundamental
dynamics (Ornstein-Uhlenbeck dynamics) reflected inside an interval
$[\underline{f},\overline{f}]$. In this section, the fundamental is
driven by the mean-reverting dynamics:
 $$
 df = \lambda (\mu - f) dt + \sigma dW_t,
 $$%
 where $\mu$ is the ``long-run'' level of the fundamental, and
 $\lambda $ is now the speed of convergence, to highlight the mean-reverting equivalent of the DMPS process.
 Following the previous exposition, we can obtain the full solution
 for the exchange rate $X^*(t, f)$ as the solution of

$$
\partial_t X + \frac{\sigma^2}{2} \partial_{ff} X + \lambda (\mu - f)
\partial_f X - \frac{\alpha}{1-r} X = - \frac{r \alpha}{1-r} f.
$$
As before, we have the stationary solution for a vanishing
$\partial_t$, and here it reads
\begin{eqnarray}
  X_S(f) &=& A \ \text{}_1F_1 \left [ \frac{\alpha}{2\lambda (1-r)} , \frac{1}{2} ; \frac{\lambda }{\sigma^2}(f - \mu)^2 \right ] +\nonumber    \\
         &+& B \frac{\sqrt{\lambda }}{\sigma}(f - \mu) \ \text{}_1F_1 \left [ \frac{\alpha}{2\lambda (1-r)} + \frac{1}{2}, \frac{3}{2} ; \frac{\lambda }{\sigma^2}(f - \mu)^2 \right] +\nonumber \\
         &+&\left [ \frac{\lambda \mu (1-r) f+ r \alpha}{\lambda (1-r) + \alpha} \right ]
\end{eqnarray}
where $\text{}_1F_1[a,b; x]$ is the confluent hypergeometric
function. The integration constants $A$ and $B$, as before, are
determined via smooth pasting at the target zone boundaries, namely:
$\partial X_S(f) |_{f = \underline{f}} = \partial X_S(f) |_{f =
  \bar{f}} = 0$.  Note that if $\mu=0$, then $A=0$. Figure
\ref{oustat} shows the stationary dynamics of the exchange rate as
function of the fundamental, for different values of long-run level
$\mu$ and noise variance $\sigma$. The band is assumed symmetric
around 0, and $\bar{f} = 10\%$.

The associated Sturm-Liouville equation is now given by

$$
\frac{\sigma^2}{2} \partial_{ff} X + \lambda (\mu - f) \partial_f X +
\rho X = 0,
$$
where $\rho = \frac{\alpha}{1-r}$, and the spectrum of the process can
be obtained explicitly by solving a transcendental equation involving
Weber parabolic cylinder functions. As before, the complete solution
is given by an expansion on a complete set of orthogonal functions on
the target band, namely:

$$
X^*(T-t, f) = X_S(f) + \sum_{k=1}^\infty c_k \exp[-(\Omega_k + \rho)
(T-t) ] \psi(\Omega_k,f),
$$
where the Fourier coefficients $c_k$ again impose the terminal
condition $X^*(0,f) = - X_S(f)$. As worked out by
\cite{linetsky2005transition} explicit though lengthy closed form
expressions are obtainable (see Eqs.(39) and (40). For the case of a
symmetric target zone $\underline{f} = - \overline{f}$, an
approximation valid for large eigenvalues $\Omega_k$, (i.e. large
$k$'s) is given in [L] and reads:
  
\begin{eqnarray}
  \label{AASY}
  \Omega_k &=& {k^{2} \pi  \sigma^{2} \over 8 \overline{f}^{2}}+ \frac{\lambda }{2} + c_0 + O\left (\frac{1}{k^2} \right ) \nonumber \\
  c_0 &=& \frac{\lambda ^2}{6 \sigma^2} (4 \bar{f}^2 - 6 \bar{f} \mu + 3 \mu^2).
\end{eqnarray}
The normalised eigenfunctions, also up to
$O\left (\frac{1}{k^2} \right )$, read:

\begin{eqnarray}
  \psi_k(f) &=& \pm \frac{\sigma}{\sqrt{2}} \bar{f}^{-1/2}\exp \left [ \frac{\lambda (f - \mu)^2}{2 \sigma^2} \right ] \left [\cos \left (\frac{k \pi f}{2\bar{f}} \right ) + \frac{2\bar{f}}{k \pi \sigma^2} \phi(f) \sin\left( \frac{k \pi f}{2 \bar{f}} \right )   \right  ] \nonumber \\
  \phi(f)  &=&  \frac{\lambda ^2}{6 \sigma^2} f^3 - \frac{\lambda ^2\mu}{2 \sigma^2} f^2 - \left [ \frac{\lambda }{2} \left ( \frac{\sqrt{2 \lambda }}{\sigma} \mu+1  + c_0 \right )\right ] f + \theta \mu
\end{eqnarray} 

%

While strictly speaking Eq.(\ref{AASY}) furnishes very good estimates
for large $k$ values, a closer look in [L] shows that even for low
$k$'s, ($k=1,2, \cdots )$, pretty good approximations are also
obtainable. In particular, for $k=1$, we approximately have:
$$
\tau _{{\rm relax}} \simeq\left[ \Omega_1\right]^{-1} = \left[ {\pi
    \sigma^{2} \over 8 \overline{f}^{2}}+ \frac{\lambda }{2} + c_0
\right]^{-1}.
$$

For this mean-reverting dynamics, the interplay between risk (here
solely due to the noise source variance $\sigma^2$) and the target
band width $2\overline{f}$ on $t_{\rm relax}$ is opposite compared to
the DMPS dynamics of section 2.


The tendency of the fundamental $f$ to revert to its long-run level
$\mu$, for a narrow target band, generates an effect of an increase in
risk (variance) that is opposite of the one generated by an increase
of $\beta$ in the DMPS setting, because of the latter's tendency to
escape from the mean. If the band is larger, lower levels of $\sigma$
initially increase the relaxation time, to ultimately achieving a
decreasing effect. In both cases, an increase in the size of the
target band requires a higher $T$ in order for the target zone to be
feasible.

We lastly notice that for the O-U case, zero is always the first
eigenvalue (not surprising, given that it's an ergodic process) and a
regime shift cannot be possible.

\section{Alternative to O-U dynamics: softly attractive drift}
\label{soft}

\noindent We now present the model where we model the fundamental as
an ergodic process with a softly attractive drift instead of the
Ornstein-Uhlenbeck dynamics. This framework has the advantage of
incorporating mean-reverting dynamics while retaining analytical
tractability. By ``softly attractive'' drift we mean the DMPS drift
with opposite sign, i.e. $-\beta\tanh(\beta f)$. This model presents
similar dynamics to the O-U framework, and allows for a stationary
time-independent probability measure. The marginal difference with the
O-U advantage is that the reversion of the fundamental to the mean is
softer, and the advantage is that the full spectrum is available and
the dynamics do not require an approximation. The equation for the
exchange rate after applying It\^o's lemma is now given by

\begin{equation}
  \label{EXD1}
  \partial_t X(t,f) + {1\over 2} \partial_{ff} X(t,f) - \beta \tanh(\beta f)  \partial_f X(t, f)  - \alpha  X(t,f) = -  \alpha   f.
\end{equation}
Using the equivalent transformation as in the DMPS case, we plug in
Eq.(\ref{EXD1}) into Eq.(\ref{DARBOUX}) and obtain:

\begin{equation}
  \label{ONESTAGE}
  \begin{array}{l}
    \int^{f} \cosh(\beta \zeta) \partial_t  Y(t, \zeta)  + \\ \\
 
    \qquad \qquad \qquad  {1 \over 2} \left[ \beta \sinh(\beta f) Y(t,f) + \cosh (\beta f) \partial_f Y(t,f) \right] - \\ \\
 
    \qquad \qquad \qquad \qquad \quad \beta \sinh(\beta f) Y(t,f) -  \alpha  \int^{f} \cosh(\beta \zeta) Y(t, \zeta) d\zeta  = -  \alpha   f. 
 
  \end{array}
\end{equation}

\noindent Now, taking once more the derivative of Eq.(\ref{ONESTAGE})
with respect to $f$, one obtains:

\begin{equation}
  \label{TWOSTAGE} 
  \partial_t Y(t, f)  + {1 \over 2} \partial_{ff}  Y(t, f)  - \left[ {\beta^{2} \over2} +  \alpha  \right] Y(t, f) = -  \alpha   {f \over \cosh(\beta f)}.
\end{equation}
Observe now that Eq.(\ref{TWOSTAGE}) is once again equivalent to the
standard BM motion case and we can repeat the same procedure we . The
spectrum will now include the eigenvalue zero since we deal with a
stationary case.

\vspace{0.5cm}
\noindent We now proceed as before and Eq.(\ref{TWOSTAGE}) reads:

\begin{equation}
  \label{TWOSTAGEBIS} 
  -\partial_{\tau} Y(\tau, f)  + {1 \over 2} \partial_{ff}  Y(\tau, f)  - \left[ {\beta^{2} \over2} +  \alpha  \right] Y(\tau, f) = -  \alpha   {f \over \cosh(\beta f)}.
\end{equation}

\noindent Consider now the homogenous part of Eq.(\ref{TWOSTAGEBIS}),
namely:

$$
\label{HOMOG}
-\partial_{\tau} Y(\tau, f) + {1 \over 2} \partial_{ff} Y(\tau, f) -
\left[ {\beta^{2} \over2} +  \alpha  \right] Y(\tau, f) =0.
$$

\noindent As done before, the method of separation of variables leads
us to introduce $Y(\tau, f) = \phi(\tau) \psi (f)$ and the previous
equation can be rewritten as:

$$
\label{SEPAR} {-\partial_{\tau} \psi (\tau) \over \psi(\tau) }+ {1
  \over 2}{ \partial_{ff} \psi(f) \over \psi(f) } - \left[ {\beta^{2}
    \over2} +  \alpha  \right]=0.
$$

\noindent and therefore we can write:

$$
\label{SEPARBIS}
\left\{
  \begin{array}{l}
    {-\partial_{\tau} \psi (\tau) \over \psi(\tau)  }=  \lambda_k, \\ \\
    {1 \over 2}{ \partial_{ff} \psi(f) \over \psi(f) } - \left[
    {\beta^{2} \over2} +  \alpha  \right] = \lambda_k
  \end{array}
\right.
$$

\noindent Defining
$\Omega^{2}_k = \left[ {\beta^{2} \over2} +  \alpha  \right]
+ \lambda_k$, the relevant eigenfunctions reads:

$$
\label{EIGEN}
\psi(f) = c_1 \sin( \sqrt{2}\Omega_k f) + c_2 \cos( \sqrt{2}\Omega_k
f).
$$

\noindent Going back to Eq.(\ref{DARBOUX}), the boundary conditions at
the borders of the target zone $\overline{f} = -\underline{f} $ reads:

$$
\label{BORDERS}
\partial_f \left[ \int^{f} \cosh(\beta \zeta) \psi(\zeta) d\zeta
\right] \mid _{f = \overline{f}}\,\, =0.
$$

\noindent which implies that:

\begin{equation}
  \label{BORDERSBIS}
  \cosh(\beta \overline{f})  \psi(\overline{f})    \qquad \Rightarrow \qquad  c_1=0 \quad {\rm and }\quad †\Omega_k  =  (2k+1) {\pi \over 2 \sqrt{2}\,\, \overline{f}}.
\end{equation}

We note that Eq.(\ref{BORDERSBIS}) implies :

\begin{equation}
  \label{CONDI}
  \lambda_k = {(2 k+1)^{2} \pi^{2} \over 8 \overline{f}^{2} } - {\beta^{2} \over 2}  -  \alpha  \geq 0.
\end{equation}
Lastly, as expected, for the soft attractive case we are able to
derive the exact spectrum analytically and unlike in Proposition 2, there is no
spectral gap.

\end{appendices}

\end{document}